\documentclass[aps,prd,preprint,showpacs,superscriptaddress,nofootinbib,preprintnumbers]{revtex4}

\usepackage{array}
\usepackage{slashed}
\usepackage{epsf,epsfig}
\usepackage{subfigure}  
\usepackage{graphicx} 
\usepackage{amsmath, amsfonts}
\usepackage{amssymb}

\allowdisplaybreaks

\def\bfnabla{\mbox{\boldmath $\nabla$}}

\def\bfsigma{\mbox{\boldmath $\sigma$}}

\def\lQ{\Lambda_{\textrm{QCD}}}
\def\al{\alpha}
\def\als{\alpha_{\textrm s}}

\def\siml{{\ \lower-1.2pt\vbox{\hbox{\rlap{$<$}\lower6pt\vbox{\hbox{$\sim$}}}}\ }} 
\def\simg{{\ \lower-1.2pt\vbox{\hbox{\rlap{$>$}\lower6pt\vbox{\hbox{$\sim$}}}}\ }}

\def\vbfD{{\ \lower-8pt\vbox{\hbox{\rlap{$\!\leftrightarrow$}\lower8pt\vbox{\hbox{$\!\bf D$}}}}\ }} 
\def\dsl{\,\raise.15ex\hbox{/}\mkern-13.5mu D}

\newcommand{\nn}{\nonumber}
\newcommand{\be}{\begin{equation}} 
\newcommand{\ee}{\end{equation}}
\newcommand{\bea}{\begin{eqnarray}} 
\newcommand{\eea}{\end{eqnarray}}
\newcommand{\beq}{\begin{equation}}
\newcommand{\eeq}{\end{equation}}
\newcommand{\bqa}{\begin{eqnarray}}
\newcommand{\eqa}{\end{eqnarray}}

\newcommand{\ka}{\sigma}

\def\lla{\langle\!\langle}
\def\rra{\rangle\!\rangle}

\newcommand{\Appendix}[1]%
    {%
     \section{#1}%
      }

\begin{document}
\preprint{\vbox{\halign{ &# \hfil\cr & TUM-EFT 36/12 \cr &\today\cr}}}

\title{Effective string theory and the long-range relativistic corrections to the quark-antiquark potential}

\author{Nora Brambilla}
\affiliation{Physik-Department, Technische Universit\"at M\"unchen,
James-Franck-Str. 1, 85748 Garching, Germany}
\author{Michael Groher}
\affiliation{Institut f\"ur Theoretische Physik, ETH Z\"urich, CH-8093 Z\"urich, Switzerland}
\author{Hector E. Martinez}
\affiliation{Physik-Department, Technische Universit\"at M\"unchen,
James-Franck-Str. 1, 85748 Garching, Germany}
\author{Antonio Vairo}
\affiliation{Physik-Department, Technische Universit\"at M\"unchen,
James-Franck-Str. 1, 85748 Garching, Germany}

\begin{abstract} 
The complete expression of the heavy quark-antiquark potential up to order $1/m^2$ is known from QCD in terms of Wilson loop expectation values.
We use that expression and a mapping, assumed to be valid at large distances, between Wilson loop expectation values 
and correlators evaluated in the effective string theory, to compute the potential. 
We obtain previously unknown results for the spin and momentum-independent parts of the potential.
These are linearly rising with the distance and may be interpreted as relativistic corrections to the string tension.
We confirm known results for the other parts of the potential. 
Finally, we compute the discrete spectrum of a heavy quark-antiquark pair 
whose interaction is just given by the obtained potential.
\end{abstract}


\maketitle
\newpage

\section{Introduction}
Wilson loops have been related to the heavy quark-antiquark potential since 
the inception of QCD~\cite{Wilson:1974sk,Susskind:1976pi,Fischler:1977yf,Brown:1979ya,Eichten:1980mw,Peskin:1983up,Barchielli:1986zs}. 
This relation has been put in a systematic framework by non-relativistic 
effective field theories of QCD~\cite{Brambilla:2000gk,Pineda:2000sz,Brambilla:2003mu,Brambilla:2004jw}.
In this framework, the heavy quark-antiquark potential is organized as an expansion in $1/m$, where $m$ is the generic heavy-quark mass, 
while non-analytic terms in $1/m$ factorize. 
Non-analytic terms may be identified with the Wilson coefficients of non-relativistic QCD (NRQCD), which is the effective field theory 
that follows from QCD by integrating out modes that scale like $m$~\cite{Caswell:1985ui,Bodwin:1994jh}.
The order $1/m^0$ potential is the static potential. It is related to the expectation value of a rectangular Wilson 
loop stretching over time and over the distance between the heavy quark and antiquark.
Contributions to the potential of higher orders in $1/m$ are expressed in terms of expectation values of chromoelectric and chromomagnetic 
field insertions on a rectangular Wilson loop. These, as well as the Wilson loop, are gauge invariant.
At order $1/m^2$, the potential is momentum and spin dependent. 

The heavy quark-antiquark potential is a function of $r$, the distance between the heavy quark and 
antiquark, and $\lQ$, the typical hadronic scale. The potential may be evaluated perturbatively for $r\lQ \ll 1$, 
but it cannot be for $r\lQ \simg 1$. The situation $r\lQ \simg 1$ is particularly relevant 
for excited charmonium and bottomonium states and for this reason has been extensively studied in lattice 
QCD~\cite{Michael:1985wf,Michael:1985rh,Campostrini:1986ki,Campostrini:1987hu,deForcrand:1985zc,Huntley:1986de,Koike:1989jf,Born:1993cp,Bali:1997am,Bali:2000gf}.
The most recent determinations are in~\cite{Koma:2006si,Koma:2006fw,Koma:2007jq,Koma:2009ws,Koma:2010zz}.
However, not all the long-range contributions to the heavy quark-antiquark potential have been computed on the lattice. 
While the order $1/m^0$ and $1/m$ contributions have been computed, as well as at order $1/m^2$ the spin and momentum-dependent potentials, 
an evaluation of the spin and momentum-independent $1/m^2$ potentials in the long range is still missing.
The reason is that they involve Wilson loops with three or four field insertions, whose lattice determination is difficult. 

The static potential measured by (quenched) lattice simulations exhibits a typical Cornell-potential type behaviour 
with a Coulombic short-range part and a linear-rising long-range tail. In the long range, $r\lQ \gg 1$, a linear potential is predicted 
by the effective string theory (EST)~\cite{Nambu:1978bd}. Long-range corrections to the linear potential have been calculated 
in the EST and confirmed by lattice simulations~\cite{Luscher:1980fr,Luscher:1980ac,Polchinski:1991ax,Luscher:2002qv}.
In~\cite{Kogut:1981gm} a one-to-one correspondence between correlators of string coordinates and field insertions 
on a rectangular Wilson loop was suggested and used to evaluate the spin-spin potential. Following that approach, 
in~\cite{PerezNadal:2008vm} the $1/m$ potential as well as all momentum and spin-dependent $1/m^2$ potentials were evaluated in the EST. 
Remarkably, in all the available cases the long-range behaviour of the (quenched) lattice data agrees with the EST determination.\footnote{ 
For the spin and momentum-dependent $1/m^2$ potentials these results were known for a long time 
in an equivalent approach to the EST that consists in approximating the Wilson loop with the exponential 
of its rectangular area~\cite{Barchielli:1986zs,Barchielli:1988zp,Brambilla:1993zw,Brambilla:1996aq,Brambilla:1999ja}. 
See also~\cite{Bali:1997am}.}
This suggests that the EST may serve to evaluate the long-range behaviour of the still unknown spin and momentum-independent
$1/m^2$ potentials, providing at the same time a non-trivial prediction for future lattice determinations 
and the missing ingredient needed to include all $1/m^2$ potentials in the computation of the quarkonium spectrum. 
The aim of this work is to address such an evaluation.

The paper is organized in the following way. In section~\ref{sec_wilson}, we establish our notation and write 
the  heavy quark-antiquark potential in terms of Wilson loop expectation values. In section~\ref{sec_EST}, 
we review the EST. In section~\ref{sec_potentials}, we derive the potential up to order $1/m^2$ in terms of EST 
correlators and in section~\ref{sec_spectrum} we look at the impact of the different parts of the $1/m^2$ potential 
on the spectrum in a model that includes only the long-range tail of the potential.
Finally, in section~\ref{sec_conclusions}, we draw some conclusions.

\section{Relativistic corrections to the static potential}
\label{sec_wilson}
The complete heavy quark-antiquark potential up to order $1/m^2$ has been written
in terms of Wilson loop expectation values in~\cite{Brambilla:2000gk,Pineda:2000sz}.
We will use here the same notations and expressions, which we recall shortly in the next two sections.

\subsection{The structure of the potential}
\label{subsecA}
We consider a heavy quark of mass $m_1$ located at ${\bf x}_1$ and a heavy antiquark 
of mass $m_2$ located at ${\bf x}_2$. The spin and momentum operators of the two particles 
are respectively ${\bf S}_1 \equiv \bfsigma_1/2$ and ${\bf p}_1 \equiv -i\bfnabla_{{\bf x}_1}$, and 
${\bf S}_2 \equiv \bfsigma_2/2$ and ${\bf p}_2 \equiv -i\bfnabla_{{\bf x}_2}$. The distance between the 
quark and the antiquark is ${\bf r} \equiv {\bf x}_1-{\bf x}_2$.
Up to order $1/m^2$ the quark-antiquark potential can be written as the sum of three terms,
\be
V = V^{(0)}  + V^{(1/m)} + V^{(1/m^2)}\,,
\ee
where $V^{(0)}(r)$ is the static potential, 
\be
V^{(1/m)}(r)  = \frac{V^{(1,0)}(r)}{m_1}+\frac{V^{(0,1)}(r)}{m_2}\,,
\ee
the $1/m$ potential and 
\be
V^{(1/m^2)} = \frac{V^{(2,0)}}{m_1^2}+\frac{V^{(0,2)}}{m_2^2}+\frac{V^{(1,1)}}{m_1 m_2}\,,
\ee
the $1/m^2$ potential.  Invariance under charge conjugation and particle interchange implies $V^{(1,0)}(r)= V^{(0,1)}(r)$. 
It is useful to separate in the  $1/m^2$ potential a spin-dependent ($SD$) from a spin-independent ($SI$) part:
\bqa
V^{(2,0)}&=&V^{(2,0)}_{SD}+V^{(2,0)}_{SI}\,,
\\
V^{(0,2)}&=&V^{(0,2)}_{SD}+V^{(0,2)}_{SI}\,,
\eqa 
where
\bqa
V_{SI}^{(2,0)}&=&\frac{1}{2}\left\{\mathbf{p}^2_1,V_{{\bf p}^2}^{(2,0)}(r)\right\}+\frac{V_{{\bf L}^2}^{(2,0)}(r)}{r^2}\mathbf{L}_1^2+V^{(2,0)}_r(r)\,,
\\
V_{SI}^{(0,2)}&=&\frac{1}{2}\left\{\mathbf{p}^2_2,V_{{\bf p}^2}^{(0,2)}(r)\right\}+\frac{V_{{\bf L}^2}^{(0,2)}(r)}{r^2}\mathbf{L}_2^2+V^{(0,2)}_r(r)\,,
\eqa 
and $\mathbf{L}_i=\mathbf{r}\times \mathbf{p}_i$ with $i=1,2$. 
Also in this case invariance under charge conjugation and particle interchange yields
\bqa
V_{{\bf p}^2}^{(2,0)}(r)&=&V_{{\bf p}^2}^{(0,2)}(r)\,,
\\
V_{{\bf L}^2}^{(2,0)}(r)&=&V_{{\bf L}^2}^{(0,2)}(r)\,,
\\
V_{r}^{(2,0)}(r)&=&V_{r}^{(0,2)}(r;m_2\leftrightarrow m_1)\,.
\eqa
For the spin-dependent part we have
\bqa
V_{SD}^{(2,0)}&=&V_{LS}^{(2,0)}(r)\,\mathbf{L}_1\cdot\mathbf{S}_1\,,
\\
V_{SD}^{(0,2)}&=&-V_{LS}^{(0,2)}(r)\,\mathbf{L}_2\cdot\mathbf{S}_2\,.
\eqa 
Charge conjugation and particle interchange invariance imply $V_{LS}^{(2,0)}(r)=V_{LS}^{(0,2)}(r;m_2\leftrightarrow m_1)$.
One proceeds similarly for the $V^{(1,1)}$ potential:
\be
V^{(1,1)}=V^{(1,1)}_{SD}+V^{(1,1)}_{SI}\,,
\ee 
where
\be
V_{SI}^{(1,1)}=-\frac{1}{2}\left\{\mathbf{p}_1\cdot\mathbf{p}_2,V_{{\bf p}^2}^{(1,1)}(r)\right\}
-\frac{V_{{\bf L}^2}^{(1,1)}(r)}{2r^2}(\mathbf{L}_1\cdot\mathbf{L}_2+\mathbf{L}_2\cdot\mathbf{L}_1)+V_r^{(1,1)}(r)\,,
\ee 
and
\be
V_{SD}^{(1,1)}=V^{(1,1)}_{L_1 S_2}(r)\mathbf{L}_1\cdot\mathbf{S}_2-V^{(1,1)}_{L_2 S_1}(r)\mathbf{L}_2\cdot\mathbf{S}_1
+V^{(1,1)}_{S^2}(r)\mathbf{S}_1\cdot\mathbf{S}_2+V^{(1,1)}_{{\bf S}_{12}}(r)\mathbf{S}_{12}(\mathbf{\hat{r}}),
\ee 
with 
\be
\mathbf{S}_{12}(\mathbf{\hat{r}}) \equiv 3\,\mathbf{\hat{r}}\cdot\bfsigma_1\,\mathbf{\hat{r}}\cdot\bfsigma_2-\bfsigma_1\cdot\bfsigma_2\,,
\ee 
and $V^{(1,1)}_{L_1 S_2}(r)=V^{(1,1)}_{L_2 S_1}(r;m_1\leftrightarrow m_2)$.

\subsection{The potential in QCD}
\label{subsecB}
In the following, we list the potentials $V^{(i,j)}(r)$ written in terms of operator insertions on a rectangular Wilson loop. 
We refer the reader to~\cite{Brambilla:2000gk,Pineda:2000sz} for the derivation of these expressions and for further details. 

The static potential is given by
\be
V^{(0)}(r)=\lim_{T\to \infty}\frac{i}{T}\ln\langle W_\Box\rangle\,,
\label{V0}
\ee 
where $\langle W_\Box\rangle$ is the expectation value of the rectangular Wilson loop,  
\be
W_\Box \equiv {\rm P} \exp\left\{{\displaystyle - i g \oint_{r\times T} \!\!dz^\mu \,A_{\mu}(z)}\right\}\,,
\label{wilson_loop}
\ee 
and P stands for the path ordering of the color matrices~\cite{Brown:1979ya}.
We also define $\langle\!\langle \dots \rangle\!\rangle\equiv \langle \dots
W_\Box\rangle / \langle W_\Box\rangle$ and the connected correlators 
\bqa
&&\lla O_1(t_1)O_2(t_2)\rra_c= \lla O_1(t_1)O_2(t_2)\rra 
-\lla O_1(t_1)\rra\lla O_{2}(t_2)\rra \,,
\label{con1}\\ 
&& \nn\\
&&\lla O_1(t_1)O_2(t_2)O_3(t_3)\rra_c= \lla O_1(t_1)O_2(t_2)O_3(t_3)\rra -\lla O_1(t_1)\rra \lla O_{2}(t_{2})O_3(t_3)\rra_c
\nn\\
& & \hspace{1.5cm}
-\lla O_1(t_1)O_2(t_2)\rra_c\lla O_{3}(t_{3})\rra
-\lla O_1(t_1)\rra \lla O_{2}(t_{2})\rra \lla O_3(t_3)\rra \,,
\label{con2}\\
&& \nn\\
&&\lla O_1(t_1)O_2(t_2)O_3(t_3)O_4(t_4)\rra_c= 
\lla O_1(t_1)O_2(t_2)O_3(t_3)O_4(t_4)\rra 
\nn\\
\nn
& &\hspace{1.5cm}
-\lla O_1(t_1)\rra \lla O_2(t_2)O_3(t_3)O_4(t_4)\rra_c
-\lla O_1(t_1)O_2(t_2)\rra_c\lla O_3(t_3)O_4(t_4)\rra_c
\\
\nn
& & \hspace{1.5cm}
-\lla O_1(t_1)O_2(t_2) O_3(t_3)\rra_c\lla O_4(t_4)\rra
-\lla O_1(t_1)\rra \lla O_2(t_2)\rra \lla O_3(t_3)O_4(t_4)\rra_c
\\
\nn
& & \hspace{1.5cm}
-\lla O_1(t_1)\rra \lla O_2(t_2)O_3(t_3)\rra_c \lla O_4(t_4)\rra
-\lla O_1(t_1)O_2(t_2)\rra_c \lla O_3(t_3)\rra \lla O_4(t_4)\rra
\\
& & \hspace{1.5cm}
-\lla O_1(t_1)\rra \lla O_2(t_2)\rra\lla O_3(t_3)\rra \lla O_4(t_4)\rra\,,
\label{con3}
\eqa 
where $O_1(t_1)$, $O_2(t_2)$, ..., $O_n(t_n)$ are operators inserted on the Wilson loop
at times $t_1 \ge t_2 \ge \dots \ge t_{n-1} \ge t_n$. Connected correlators are made of 
Feynman diagrams that cannot be disconnected by cutting once the heavy-quark and antiquark lines.

The $1/m$ potential is given by 
\be
V^{(1,0)}(r) = - \frac{1}{2} \int_0^\infty dt \, t \, \lla g{\bf E}_1(t)\cdot g{\bf E}_1(0) \rra_c\,,
\ee
where ${\bf E}_i(t)$ (and later  ${\bf B}_i(t)$) stands for ${\bf E}(t,{\bf x}_i)$ (${\bf B}(t,{\bf x}_i)$) with $i=1,2$.
The $1/m^2$ potentials are\footnote{
We have dropped terms proportional to $\bfnabla_r^i V^{(0)}$ in the expressions of $V_r^{(2,0)}(r)$ and $V_r^{(1,1)}(r)$ 
because they are suppressed in the non-relativistic power counting (see section VI of~\cite{Pineda:2000sz}). 
} 
\bqa
V_{{\bf p}^2}^{(2,0)}(r) &=& \frac{i}{2}{\hat {\bf r}}^i{\hat {\bf r}}^j
\int_0^\infty dt \,t^2 \lla g{\bf E}_1^i(t) g{\bf E}_1^j(0) \rra_c\,,
\\
V_{{\bf L}^2}^{(2,0)}(r) &=& \frac{i}{4}
\left(\delta^{ij}-3{\hat {\bf r}}^i{\hat {\bf r}}^j \right)
\int_0^\infty dt \, t^2 \lla g{\bf E}_1^i(t) g{\bf E}_1^j(0) \rra_c\,,
\\
V_{LS}^{(2,0)}(r) &=& - \frac{c_F^{(1)}}{r^2}i {\bf r}\cdot \int_0^\infty dt \, t \,  \lla g{\bf B}_1(t) \times g{\bf E}_1 (0) \rra 
+ \frac{c_S^{(1)}}{2 r^2}{\bf r}\cdot \left(\bfnabla_r V^{(0)}\right)\,,
\\
V_{{\bf p}^2}^{(1,1)}(r)&=& i {\hat {\bf r}}^i{\hat {\bf r}}^j
\int_0^\infty dt \,t^2 \lla g{\bf E}_1^i(t) g{\bf E}_2^j(0) \rra_c\,,
\\
V_{{\bf L}^2}^{(1,1)}(r)&=& \frac{i}{2}
\left(\delta^{ij}-3{\hat {\bf r}}^i{\hat {\bf r}}^j \right)
\int_0^\infty dt \, t^2 \lla g{\bf E}_1^i(t) g{\bf E}_2^j(0) \rra_c\,,
\\
V_{L_2S_1}^{(1,1)}(r)&=& - \frac{c_F^{(1)}}{r^2}i {\bf r}\cdot \int_0^\infty dt \, t \, 
\lla g{\bf B}_1(t) \times g{\bf E}_2 (0) \rra\,,
\\
V_{S^2}^{(1,1)}(r)&=& \frac{2 c_F^{(1)} c_F^{(2)}}{3}i \int_0^\infty  dt \,  
\lla g{\bf B}_1(t) \cdot g{\bf B}_2 (0) \rra - 4(d_{sv} + d_{vv} C_f ) \,\delta^{(3)}({\bf r})\,,
\label{spinspinQCD}
\\
V_{{\bf S}_{12}}^{(1,1)}(r)&=& \frac{c_F^{(1)} c_F^{(2)}}{4}i {\hat {\bf r}}^i{\hat {\bf r}}^j
\int_0^\infty  dt \, 
\left[
\lla g {\bf B}^i_1(t) g {\bf B}^j_2 (0) \rra  - \frac{\delta^{ij}}{3}\lla g{\bf B}_1(t)\cdot g{\bf B}_2 (0) \rra
\right]\,,
\label{tensorspinQCD}
\\
V_r^{(2,0)}(r) &=&  \frac{\pi C_f \als c_D^{(1)\prime}}{2} \delta^{(3)}({\bf r})
\nn\\
\nn
&&
- \frac{i c_F^{(1)\,2}}{4} \int_0^\infty dt \lla g{\bf B}_1(t)\cdot g{\bf B}_1(0) \rra_c
+ \frac{1}{2}\left(\bfnabla_r^2 V_{{\bf p}^2}^{(2,0)}\right)
\\
\nn
&&
- \frac{i}{2}\int_0^\infty dt_1\int_0^{t_1} dt_2 \int_0^{t_2}
dt_3\, (t_2-t_3)^2 \lla g{\bf  E}_1(t_1)\cdot g{\bf E}_1(t_2) g{\bf E}_1(t_3)\cdot g{\bf E}_1(0) \rra_c 
\\
\nn
&&
+ \frac{1}{2}
\left(\bfnabla_r^i
\int_0^\infty dt_1\int_0^{t_1} dt_2 \, (t_1-t_2)^2 \lla
g{\bf E}_1^i(t_1) g{\bf E}_1(t_2)\cdot g{\bf E}_1(0) \rra_c
\right)
\\
& & 
- d_3^{(1)\prime} f_{abc} \int d^3{\bf x} \, 
\lim_{T\rightarrow \infty} g \lla F^a_{\mu\nu}({x}) F^b_{\mu\al}({x}) F^c_{\nu\al}({x}) \rra  
\,,
\\
V_r^{(1,1)}(r) &=& -\frac{1}{2}\left(\bfnabla_r^2 V_{{\bf p}^2}^{(1,1)}\right)
\nn \\
\nn
&&
-i\int_0^\infty dt_1\int_0^{t_1} dt_2 \int_0^{t_2}
dt_3\, (t_2-t_3)^2 \lla g{\bf
  E}_1(t_1)\cdot g{\bf E}_1(t_2) g{\bf E}_2(t_3)\cdot g{\bf E}_2(0) \rra_c 
\\
\nn
&&
+
\frac{1}{2}
\left(\bfnabla_r^i
\int_0^\infty dt_1\int_0^{t_1} dt_2 (t_1-t_2)^2 
\lla g{\bf E}_1^i(t_1) g{\bf E}_2(t_2)\cdot g{\bf E}_2(0) \rra_c \right)
\\
\nn
&&
+ \frac{1}{2}
\left(\bfnabla_r^i
\int_0^\infty dt_1\int_0^{t_1} dt_2 (t_1-t_2)^2 
\lla g{\bf E}_2^i(t_1) g{\bf E}_1(t_2)\cdot g{\bf E}_1(0) \rra_c
\right)
\\
&&
+ (d_{ss} + d_{vs} C_f) \,\delta^{(3)}({\bf r}) 
\,.
\eqa 
The coefficients $c_F^{(i)} = 1 + {\cal O}(\als)$, $c_S^{(i)} = 2 c_F^{(i)}-1$, $c_D^{(i)\prime} = 1 + {\cal O}(\als)$, 
$d_3^{(1)\prime} = \als/(720\pi) + {\cal O}(\als^2)$~\cite{Manohar:1997qy}, and $d_{sv}$, $d_{vv}$, 
$d_{ss}$, $d_{vs}$, which are such that $(d_{sv} + d_{vv} C_f) = {\cal O}(\als^2)$ and 
$(d_{ss} + d_{vs} C_f)  = {\cal O}(\als^2)$~\cite{Pineda:1998kj}, are Wilson coefficients of NRQCD. 
The natural scale of $\als$ in these coefficients is of the order of the heavy-quark mass, 
hence we may expect $\als$ to be a fairly small number.
The constant $C_f$ is the Casimir of the fundamental representation of SU(3): $C_f = 4/3$.

\section{The Effective string theory}
\label{sec_EST}
The effective string theory hypothesis states that in pure gluodynamics and in the long-distance regime, $r\lQ \gg 1$, 
the expectation value of the rectangular Wilson loop can be given in terms of a string action:
\begin{equation} 
\lim_{T\to\infty} \langle W_\Box \rangle = Z \int \mathcal{D} \xi^1 \mathcal{D} \xi^2 \, e^{iS_{\rm string}(\xi^1,\xi^2)} \,,
\label{eq:translation} 
\end{equation}
where $Z$ is a constant.\footnote{
For a general discussion about our current understanding of the QCD vacuum as it is obtained from lattice gauge theory 
and the duality to string theory we refer to~\cite{Brambilla:2014jmp}. 
For recent developments on the effective theory of long strings we refer to~\cite{Aharony:2010cx,Aharony:2013ipa}.
The effective string theory may also provide a long-distance description for other models, 
an example being the Abrikosov–-Nielsen–-Olesen vortices of the abelian Higgs model~\cite{Baker:2000ci,Baker:2002km}.
} 
The string action, $S_{\rm string}$, can be expanded in a series whose terms involve an increasing number 
of derivatives acting on the transverse string coordinates $\xi^l = \xi^l(t,z)$ ($l=1,2$)~\cite{Luscher:1980ac}. 
The coordinates $\xi^l$ count like $1/\lQ$, whereas derivatives in $t$ and $z$ acting on them count like $1/r$.
Hence, terms in $S_{\rm string}$ with more derivatives are suppressed in the long range by powers of $1/(r\lQ)$
with respect to terms with less derivatives. Up to terms with only two derivatives, the string action reads
\begin{equation} 
S_{\rm string} = -\sigma \int dt \, dz \; \left(1-\frac{1}{2} \partial_\mu \xi^{l} \partial^{\mu} \xi^{l}\right). 
\label{eq:sstring} 
\end{equation}
Studies constraining the form of the higher-order terms, also by Lorentz invariance, are
in~\cite{Luscher:2004ib,Aharony:2010cx,Aharony:2013ipa}.
The first next terms in the expansion turn out to involve at least four
derivatives and are suppressed by  $1/(r\lQ)^2$ with respect to the kinetic term in \eqref{eq:sstring}.
Such terms and subleading ones do not affect the results presented in this work and will be neglected in the rest of the paper.
Since the string has fixed ends at $z=-r/2$ and $z=r/2$, the transverse coordinates $\xi^l$ satisfy the boundary conditions $\xi^l(t,-r/2) = \xi^l(t,r/2) = 0$. 
The constant $\sigma$, which is of order $\lQ^2$, is the string tension. 
Its numerical value is known from lattice QCD determinations. 
From \eqref{V0}, \eqref{eq:translation} and \eqref{eq:sstring} it follows that~\cite{Luscher:1980fr,Luscher:1980ac}
\be
V^{(0)}(r)=\sigma r +\mu -\frac{\pi}{12r} \approx \sigma r \,,
\label{v0est}
\ee 
where $\mu$ is an unknown regularization-dependent constant 
and the term $-\pi/(12 r)$ is a universal quantum correction known as the L\"uscher term.\footnote{
The L\"uscher term does depend on the dimension of space-time. In $d$ dimensions it reads
$-\pi(d-2)/(24\,r)$. Equation \eqref{v0est} holds for $d=4$. 
} 
The last approximation holds in the large distance limit when the L\"uscher term may be neglected.

In~\cite{Kogut:1981gm} it was proposed that the mapping~\eqref{eq:translation} could be extended  
to relate Wilson loops with field strength tensor insertions to correlators of the string fields~$\xi^l$.
This would allow to compute in the EST the long-range tail of the potentials listed in section~\ref{subsecB}: 
a program started with~\cite{Kogut:1981gm} and expanded in~\cite{PerezNadal:2008vm}. 
We will follow this latter reference.
Requiring the same symmetry properties for the transverse string coordinates 
and the operators inserted in the Wilson loop, the following
mapping between expectation values of operators inserted in the Wilson loop  
and EST correlators can be established for  $r\lQ \gg 1$:
\begin{align}
\lla \dots \mathbf{E}_1^{l}(t) \dots \rra & = 
\langle \dots \Lambda^2 \partial_z \xi^l (t,{r}/{2}) \dots \rangle\,,
\nonumber \\
\lla \dots \mathbf{E}_2^{l}(t) \dots \rra  & = 
\langle \dots \Lambda^2 \partial_z \xi^l (t,-{r}/{2}) \dots \rangle\,,
\nonumber \\
\lla \dots \mathbf{B}_1^{l}(t) \dots \rra  & = 
\langle \dots \Lambda' \epsilon^{lm} \partial _t \partial_z \xi^m (t,{r}/{2}) \dots \rangle\,, 
\nonumber \\
\lla \dots \mathbf{B}_2^{l}(t) \dots \rra  & = 
\langle \dots\,\!  - \Lambda' \epsilon^{lm} \partial _t \partial_z \xi^m (t,-{r}/{2}) \dots \rangle\,, 
\nonumber \\
\lla \dots \mathbf{E}_1^{3}(t) \dots \rra  & = 
\langle \dots \Lambda''^{\,2} \dots \rangle\,,
\label{mapping} \\
\lla \dots \mathbf{E}_2^{3}(t) \dots \rra  & = 
\langle \dots  \Lambda''^{\,2} \dots \rangle\,,
\nonumber \\
\lla \dots \mathbf{B}_1^{3}(t) \dots \rra  & = 
\langle \dots \Lambda''' \epsilon^{lm} \partial _t \partial_z \xi^l (t,{r}/{2}) \partial_z \xi^m (t,{r}/{2}) \dots \rangle\,, 
\nonumber \\
\lla \dots \mathbf{B}_2^{3}(t) \dots \rra  & = 
\langle \dots\,\! - \Lambda''' \epsilon^{lm} \partial _t \partial_z \xi^l (t,-{r}/{2}) \partial_z \xi^m (t,-{r}/{2}) \dots \rangle\,, 
\nonumber
\end{align} 
where the indices $l$ and $m$ label the transverse coordinates: $l,m=1,2$. 
The tensor $\epsilon^{lm}$ is such that $\epsilon^{12}=1$ and  $\epsilon^{lm} = -\epsilon^{ml}$.
In the Wilson-loop part of the mapping the heavy quark is located at ${\bf x}_1 = (0,0,r/2)$ and 
the heavy antiquark at ${\bf x}_2 = (0,0,-r/2)$, which implies ${\bf r} = (0,0,r)$. 
The constants $\Lambda$, $\Lambda'$, $\Lambda''$ and $\Lambda'''$ are unknown constants of mass dimension one and of order $\lQ$.
The mapping \eqref{mapping} is valid up to corrections that are subleading in the long range in the EST counting.
For the purpose of the computation in this paper we will assume the mapping to be exact and neglect subleading corrections.
We will comment on the impact of subleading corrections at the end of the next section and in the conclusions.

The right-hand side of \eqref{mapping} is made of correlators of string coordinates~$\xi^l$.
The  functional integral over the string coordinates is Gaussian (see the string action \eqref{eq:sstring}). 
So we have that correlators of more than two string fields~$\xi^l$ break up into products of two-field correlators and derivatives of them, 
and that two-field correlators are given by~\cite{PerezNadal:2008vm}
\be
\langle \xi^l(it,z) \xi^m(it',z')\rangle =\frac{\delta^{lm}}{4 \pi \sigma} \ 
\ln \left(\frac{\cosh[(t-t')\,\pi/r]+\cos[(z+z')\,\pi/r]}{\cosh[(t-t')\,\pi/r]-\cos[(z-z')\,\pi/r]} \right)\,.
\label{finalcorrelator} 
\ee 
The calculation of the different possible right-hand sides of \eqref{mapping} through \eqref{finalcorrelator} 
leads to the EST long-range estimate of the potentials listed in section~\ref{subsecB}.

\section{The long-range potential in the effective string theory}
\label{sec_potentials}
The mapping \eqref{mapping} allows us to evaluate in the long range the Wilson loop expectation values that appear 
in section~\ref{subsecB}. Correlators of two string fields are given in \eqref{finalcorrelator}. 
Derivatives of two field correlators follow from it straightforwardly. 
Correlators involving more than two string fields, which come from mapping Wilson loops with ${\bf B}^3$ fields   
or more than two chromoelectric field insertions, decompose into the product of two string field correlators 
due to the Gaussian string action. Gaussianity also implies that correlators with an odd number of string fields vanish. 
Hence the Wilson loop expectation values of section~\ref{subsecB} map for $r\lQ \gg 1$ into the following expressions:
\bqa
\lla {\bf E}_1^i(it) {\bf E}_1^j (0) \rra_c &=& 
\tilde\delta^{ij} \frac{\pi\Lambda^4}{4\sigma r^2} \sinh^{-2}\left(\frac{\pi t}{2 r}\right),
\label{map1}\\
\lla {\bf E}_1^i(it) {\bf E}_2^j (0) \rra_c &=& 
- \tilde\delta^{ij} \frac{\pi\Lambda^4}{4\sigma r^2} \cosh^{-2}\left(\frac{\pi t}{2 r}\right),
\label{map2}\\
{\bf r} \cdot \lla {\bf B}_1(it) \times {\bf E}_1(0) \rra &=& 
\frac{i\pi^2\Lambda^2\Lambda^\prime}{2\sigma r^2} \cosh\left(\frac{\pi t}{2 r}\right) \sinh^{-3}\left(\frac{\pi t}{2 r}\right),
\label{map3}\\
{\bf r} \cdot \lla {\bf B}_1(it) \times {\bf E}_2(0) \rra &=& 
-\frac{i\pi^2\Lambda^2\Lambda^\prime}{2\sigma r^2} \sinh\left(\frac{\pi t}{2 r}\right) \cosh^{-3}\left(\frac{\pi t}{2 r}\right),
\label{map4}\\
\sum_{l=1}^2 \lla {\bf B}_1^l(it) {\bf B}_1^l(0) \rra_c &=&
\frac{\pi^3\Lambda'^{\,2}}{4\sigma r^4} \sinh^{-4}\left(\frac{\pi t}{2r}\right) \left[2+\cosh\left(\frac{\pi t}{r}\right)\right],
\label{map5}\\
\sum_{l=1}^2 \lla {\bf B}_1^l(it) {\bf B}_2^l(0) \rra_c &=&
-\frac{\pi^3\Lambda'^{\,2}}{4\sigma r^4} \cosh^{-4}\left(\frac{\pi t}{2r}\right) \left[2-\cosh\left(\frac{\pi t}{r}\right)\right],
\label{map6}\\
\lla {\bf B}_1^3(it) {\bf B}_1^3(0) \rra_c &=&
\frac{\pi^4\Lambda'''^{\,2}}{16\sigma^2 r^6}\sinh^{-6}\left(\frac{\pi t}{2r}\right),
\label{map7}\\
\lla {\bf B}_1^3(it) {\bf B}_2^3(0) \rra_c &=&
\frac{\pi^4\Lambda'''^{\,2}}{16\sigma^2 r^6}\cosh^{-6}\left(\frac{\pi t}{2r}\right),
\label{map8}\\
\lla {\bf E}_1(it_1)\cdot {\bf E}_1(it_2) {\bf E}_1(it_3)\cdot {\bf E}_1(0) \rra_c&=& 
\frac{\pi^2\Lambda^8}{8\sigma^2 r^4}
\left[\sinh^{-2}\left(\frac{\pi t_2}{2r}\right) \sinh^{-2}\left(\frac{\pi (t_1-t_3)}{2r}\right) \right.
\nonumber\\
&& \hspace{1.0cm} \left.
+ \sinh^{-2}\left(\frac{\pi t_1}{2r}\right) \sinh^{-2}\left(\frac{\pi (t_2-t_3)}{2r}\right)\right],
\label{map9}\\
\lla {\bf E}_1(it_1)\cdot {\bf E}_1(it_2) {\bf E}_2(it_3)\cdot {\bf E}_2(0) \rra_c&=& 
\frac{\pi^2\Lambda^8}{8\sigma^2 r^4}
\left[\cosh^{-2}\left(\frac{\pi t_2}{2r}\right) \cosh^{-2}\left(\frac{\pi (t_1-t_3)}{2r}\right) \right.
\nonumber\\
&& \hspace{1.0cm} \left.
+ \cosh^{-2}\left(\frac{\pi t_1}{2r}\right) \cosh^{-2}\left(\frac{\pi (t_2-t_3)}{2r}\right)\right],
\label{map10}
\eqa 
where $\tilde\delta^{ij} =0$ for $i$ or $j = 3$ and $\tilde\delta^{ij} = \delta^{ij}$ for $i,j = 1,2$.
The expressions for the Wilson loop expectation values with two chromomagnetic or chromoelectric
field insertions agree with those in~\cite{Kogut:1981gm}. 
Terms of the type $\lla {\bf E}^i(t_1) {\bf E}(t_2) \cdot {\bf E}(0) \rra_c$ vanish after \eqref{mapping}
regardless of the quark line where the chromoelectric fields are located. 
This is due to Gaussianity and to the subtraction of the disconnected parts; see~\eqref{con2}.\footnote{
It is also a specific feature of $\lla {\bf E}^i(t_1) {\bf E}(t_2) \cdot {\bf E}(0) \rra_c$,
which is the only type of three-field correlator appearing in the heavy quark-antiquark potential up to order $1/m^2$. 
For example, a term like $\lla {\bf E}^j(t_1) {\bf E}^3(t_2) {\bf E}^j(0) \rra_c$ would not vanish after \eqref{mapping}.}
Terms involving four chromoelectric fields contribute in the EST through 
diagrams made of two two-field correlators that are connected.

Substituting \eqref{map1}-\eqref{map10} in the expressions of the potentials, we obtain
\bqa
V^{(1,0)}(r) &=& \frac{g^2\Lambda^4}{2\pi\sigma}\ln\left(\sigma r^2\right) + \mu_1\,,
\label{v1est}\\
V^{(2,0)}_{{\bf p}^2}(r) &=& 0\,,
\label{v20p2est}\\
V^{(2,0)}_{{\bf L}^2}(r) &=& -\frac{g^2\Lambda^4\,r}{6\ka}\,,
\label{v20l2est}\\
V^{(2,0)}_{LS}(r) &=& -\frac{\mu_2}{r} -\frac{c_F^{(1)}g^2\Lambda^2\Lambda^{'}}{\ka\,r^2}\,,
\label{v20lsest}\\
V^{(1,1)}_{{\bf p}^2}(r) &=& 0\,,
\label{v11p2est}\\
V^{(1,1)}_{{\bf L}^2}(r) &=&\frac{g^2\Lambda^4\,r}{6\ka}\,,
\label{v11l2est}\\
V^{(1,1)}_{L_2\,S_1}(r) &=& -\frac{c_F^{(1)}g^2\Lambda^2\Lambda^{'}}{\ka\,r^2}\,,
\label{v11lsest}\\
V^{(1,1)}_{S^2}(r) &=& \frac{2\pi^3c_F^{(1)}c_F^{(2)}g^2\Lambda'''^{\,2}}{45\ka^2\,r^5}
-4(d_{sv}+d_{vv}C_f)\delta^{(3)}({\bf r})\,,
\label{v11s2est}\\
V^{(1,1)}_{{\bf S}_{12}}(r)&=&  \frac{\pi^3c_F^{(1)}c_F^{(2)}g^2\Lambda'''^{\,2}}{90\ka^2\,r^5}\,,
\label{v11s12est}\\
V_r^{(2,0)}(r) &=& 
-\frac{2\,\zeta_3\,g^4\Lambda^8r}{\pi^3\ka^2} +\mu_3 +\frac{\mu_4}{r^2} +\frac{\mu_5}{r^4}
+\frac{\pi^3 c_F^{(1)\,2} g^2 \Lambda'''^{\,2}}{60\sigma^2 r^5}  
\nonumber\\
&& + \frac{\pi C_f \als c_D^{(1)\prime}}{2} \delta^{(3)}({\bf r})
- d_3^{(1)\prime} f_{abc} \int d^3{\bf x} \, \lim_{T\rightarrow \infty} g \lla F^a_{\mu\nu}({x}) F^b_{\mu\al}({x}) F^c_{\nu\al}({x}) \rra  
\,,
\label{v20rest}\\
V_r^{(1,1)}(r) &=& -\frac{\zeta_3\,g^4\Lambda^8r}{2\pi^3\ka^2}
+ (d_{ss} + d_{vs} C_f) \,\delta^{(3)}({\bf r}) 
\,,
\label{v11rest}
\eqa 
where $\zeta_3 = 1.2020569...$ is the Riemann zeta function of argument three\footnote{
It comes from the integrals
\bea
\int_0^\infty  dt_1  \int_0^{t_1}  dt_2  \int_0^{t_2}  dt_3\, (t_2-t_3)^2 
\left[ \sinh^{-2}t_2\,\sinh^{-2}(t_1-t_3) + \sinh^{-2}t_1\,\sinh^{-2}(t_2-t_3) \right]
&=& 
\nonumber\\
8 \int_0^\infty  dt_1  \int_0^{t_1}  dt_2  \int_0^{t_2}  dt_3\, (t_2-t_3)^2 
\left[ \cosh^{-2}t_2\,\cosh^{-2}(t_1-t_3) + \cosh^{-2}t_1\,\cosh^{-2}(t_2-t_3) \right]
&=& \zeta_3\,.
\nonumber
\eea
} and $\mu_i$ are renormalization constants.
The expressions for the potentials \eqref{v1est}-\eqref{v11lsest} agree with those in~\cite{PerezNadal:2008vm}. 
The spin-spin potentials \eqref{v11s2est} and \eqref{v11s12est} are of order $1/r^5$.
The $1/r^5$ behaviour comes from the subleading correlator \eqref{map8}, 
for the long-range leading contribution coming from the correlator \eqref{map6}, 
which would be of order $1/r^3$, vanishes in the integrals of \eqref{spinspinQCD} and \eqref{tensorspinQCD}
(the result is independent on the specific form of the string action). 
This contrasts with the result of~\cite{Kogut:1981gm}, where the correlator \eqref{map8} is not taken into account 
and the leading spin-spin potentials shows up only at order $1/m^4$.\footnote{
The behaviour of the spin-spin potentials and the disagreement with~\cite{Kogut:1981gm} has been
pointed out in~\cite{PerezNadal:2008vm}. We thank Joan Soto for addressing our attention to this point.
}
The explicit expressions \eqref{v11s2est} and \eqref{v11s12est} are new.
The potentials \eqref{v20rest} and \eqref{v11rest} are also new.
We observe that correlators of two chromoelectric fields contracted with ${\bf r} = (0,0,r)$ 
vanish because of ${\bf r}^i\tilde\delta^{ij} = 0$, and that we do not have a mapping prescription 
into the EST for the matrix element $\lla F^a_{\mu\nu}({x}) F^b_{\mu\al}({x}) F^c_{\nu\al}({x}) \rra$
involving three gluon fields located at an arbitrary point $x$ of space-time.
The expressions listed here correct some of the preliminary findings reported in~\cite{Martinez:2012rra}.

As pointed out in~\cite{PerezNadal:2008vm}, Poincar\'e invariance fixes some of the 
renormalization constants $\mu_i$ and field normalization constants, $\Lambda$, $\Lambda'$, ..., 
because it requires some equations to be exactly fulfilled by the potentials (see~\cite{Brambilla:2001xk,Brambilla:2003nt}).
One of these equations is the Gromes relation that relates the spin-orbit potentials 
with the static potential~\cite{Gromes:1984ma}:
\be
\frac{1}{2r} \frac{dV^{(0)}}{dr} + V_{LS}^{(2,0)}-V_{L_2S_1}^{(1,1)} = 0 \,.
\label{gromes}
\ee
This equation is fulfilled in the EST only if
\be
\mu_2 = \frac{\sigma}{2} \,.
\label{mu2}
\ee
Another equation relates the momentum-dependent potentials with the static potential~\cite{Barchielli:1988zp}:\footnote{
In~\cite{Barchielli:1988zp,Brambilla:2001xk} also the exact relation
$$
-4V^{(2,0)}_{{\bf p}^2} + 2V^{(1,1)}_{{\bf p}^2}-V^{(0)}+r \frac{dV^{(0)}}{dr} =0\,,
$$
was derived. This relation is automatically fulfilled by the potentials \eqref{v0est}, \eqref{v20p2est} and \eqref{v11p2est} 
in the long range, i.e., neglecting $\mu$ and the L\"uscher term in $V^{(0)}$, and does not provide further constraints.
}
\be
\frac{r}{2} \frac{dV^{(0)}}{dr} + 2V^{(2,0)}_{\mathbf{L}^2}- V^{(1,1)}_{\mathbf{L}^2} = 0\,.
\label{BBMP}
\ee
This equation is fulfilled in the EST only if 
\be 
g\Lambda^2 = \sigma\,.
\label{Lambda}
\ee  
A similar relation holds for $\Lambda''$ and follows from the equation 
$-\bfnabla_1  V^{(0)} = \lla g {\bf E}_1\rra $ valid for $T\to\infty$ derived 
in~\cite{Brambilla:2000gk}. The equation is fulfilled in the EST only if 
\be 
g\Lambda''^{\,2} = - \sigma\,.
\label{Lambdatwo}
\ee  
Equations \eqref{Lambda} and \eqref{Lambdatwo} are remarkable, 
for they completely determine the long-range mapping of the chromoelectric field in the~EST. 
Finally, we note that the equations induced by Poincar\'e invariance would require the inclusion of subleading corrections 
to the action~\eqref{eq:sstring} and the mapping~\eqref{mapping} in order to be fulfilled beyond leading order in the long-range limit.

Taking the potentials \eqref{v1est}-\eqref{v11rest} at leading order in the long-range limit,
using the constraints \eqref{mu2} and \eqref{Lambda}, and dropping terms suppressed by powers of $\als$, 
like the term proportional to  $\lla F^a_{\mu\nu}({x}) F^b_{\mu\al}({x}) F^c_{\nu\al}({x}) \rra$, we obtain
\bqa
V^{(1,0)}(r) &=& \frac{\sigma}{2\pi}\ln\left(\sigma r^2\right)  + \mu_1\,,
\label{v1estlong}\\
V^{(2,0)}_{{\bf p}^2}(r) &=& 0\,,
\label{v20p2estlong}\\
V^{(2,0)}_{{\bf L}^2}(r) &=& -\frac{\sigma\,r}{6}\,,
\label{v20l2estlong}\\
V^{(2,0)}_{LS}(r) &=& -\frac{\sigma}{2\,r} - \frac{c_F^{(1)}g\Lambda^{'}}{r^2}\,,
\label{v20lsestlong}\\
V^{(1,1)}_{{\bf p}^2}(r) &=& 0\,,
\label{v11p2estlong}\\
V^{(1,1)}_{{\bf L}^2}(r) &=&\frac{\sigma\,r}{6}\,,
\label{v11l2estlong}\\
V^{(1,1)}_{L_2\,S_1}(r) &=&  -\frac{c_F^{(1)}g\Lambda^{'}}{r^2}\,,
\label{v11lsestlong}\\
V^{(1,1)}_{S^2}(r) &=& \frac{2\pi^3 c_F^{(1)}c_F^{(2)}g^2\Lambda'''^{\,2}}{45\ka^2\,r^5}\,,
\label{v11s2estlong}\\
V^{(1,1)}_{{\bf S}_{12}}(r)&=&  \frac{\pi^3c_F^{(1)}c_F^{(2)}g^2\Lambda'''^{\,2}}{90\ka^2\,r^5}\,,
\label{v11s12estlong}\\
V_r^{(2,0)}(r) &=& -\frac{2\,\zeta_3\,\sigma^2r}{\pi^3}\,,
\label{v20restlong}\\
V_r^{(1,1)}(r) &=& -\frac{\zeta_3\,\sigma^2 r}{2\pi^3}\,.
\label{v11restlong}
\eqa 
We have kept the subleading term proportional to $1/r^2$ in \eqref{v20lsestlong}, because 
\eqref{v0est} and \eqref{v11lsestlong} together with \eqref{gromes} guarantee that there cannot 
be any other term proportional to $1/r^2$ contributing to $V^{(2,0)}_{LS}$.
Equations \eqref{v1estlong}-\eqref{v11restlong} provide the EST expressions for the heavy quark-antiquark potential 
in the long range following from the exact mapping \eqref{mapping}. 
Power-counting arguments imply that subleading corrections to the mapping 
will not change the functional dependence of the potential but may affect some of the numerical coefficients.
This can be the case for the spin-spin potentials, which at order $1/r^5$ may be affected by subleading 
contributions proportional to two string fields in the mapping of ${\bf B}^l$, 
and for the potentials $V_r^{(2,0)}(r)$ and $V_r^{(1,1)}(r)$, 
which at order $r$ may be affected by subleading contributions proportional to two string fields in the mapping of ${\bf E}^3$.
In this last case, we note that all terms proportional to $\Lambda''^{\,8}r^5$, $\Lambda''^{\,6}r^3$ and 
$\Lambda''^{\,4}\,(\Lambda^{4}/\sigma)\,r^3$ vanish after subtraction of the disconnected parts of the correlators.

\section{Spectrum}
\label{sec_spectrum}
In order to illustrate the impact on the spectrum of the new long-range potentials derived in the previous section,  
we consider the following model: a quark-antiquark pair both of mass $m$ 
bound by the potential given in \eqref{v1estlong}-\eqref{v11restlong}.
In the centre-of-mass frame, the Hamiltonian of the system is $H = {\bf p}^2/m + V$.
The potential, $V$, reads 
\bea
V(r) &=& V^{(0)}(r) + \frac{2}{m}V^{(1,0)}(r)
+\frac{1}{m^2}\Bigg\{\left[2\frac{V_{{\bf L}^2}^{(2,0)}(r)}{r^2}+\frac{V_{{\bf L}^2}^{(1,1)}(r)}{r^2}\right] \mathbf{L}^2
\nonumber\\
&& 
+\left[V_{LS}^{(2,0)}(r)+V^{(1,1)}_{L_2\,S_1}(r)\right]\mathbf{L}\cdot\mathbf{S}
+V^{(1,1)}_{S^2}(r)\left(\frac{\mathbf{S}^2}{2}-\frac{3}{4} \right)
+V^{(1,1)}_{{\bf S}_{12}}(r)\mathbf{S}_{12}(\mathbf{\hat{r}})
\nonumber\\
&& 
+2V_{r}^{(2,0)}(r)+V_{r}^{(1,1)}(r)\Bigg\}
\nonumber\\
&\approx&\sigma r + \frac{1}{m} \frac{\sigma}{\pi}\ln\left(\sigma r^2\right)
+\frac{1}{m^2}\left(-\frac{\sigma}{6r}\mathbf{L}^2 -\frac{\sigma}{2r}\mathbf{L}\cdot\mathbf{S}-\frac{9\,\zeta_3\,\sigma^2r}{2\pi^3}\right)\,,
\label{model}
\eea
where $\mathbf{L} = {\bf r}\times {\bf p}$ and $\mathbf{S}$ is the total spin of the system. 
In the last line we have dropped contributions to the static and spin-orbit potentials that are 
subleading in the long range, and the spin-spin potentials, which fall off sharply like $1/r^5$.
The constants in the static and $1/m$ potentials do not contribute to the energy level splittings; hence we do not display them.
The model has the advantage of depending only on two parameters: the mass $m$ and the string tension $\sigma$.

We compute the energy levels by including contributions from the potential that are first order in $1/m^2$ and up to second order in $1/m$.
We call $E^{(0)}_{nl}$ the eigenvalues of the zeroth-order Hamiltonian ${\bf p}^2/m + \sigma r$.
The eigenstates of the zeroth-order Hamiltonian, $|nljs\rangle$, may be chosen to be simultaneously 
eigenstates of the angular momenta and spin. They are labeled by $n$, $l$, $j$ and $s$, which are the principal, 
orbital angular momentum, total angular momentum and spin quantum numbers. 
The state $|nl\rangle$ stands for $|nljs\rangle$ when acting on an operator that does not depend on spin.
The energy levels read\footnote{
Kinetic energy, static potential and $E^{(0)}_{nl}$ are related by the virial theorem:
$$
\langle nl | \frac{{\bf p}^2}{m} |nl\rangle = \frac{1}{2} \langle nl | \sigma r  |nl\rangle = \frac{E^{(0)}_{nl}}{3}
\sim \frac{\sigma^{2/3}}{m^{1/3}}\,,
$$
where the last relation shows the dependence of $E^{(0)}_{nl}$ on the parameters $m$ and $\sigma$~\cite{Lucha:1991vn}.
From this it follows that $1/\langle nl |r|nl \rangle \sim ( \sigma m)^{1/3}$. 
One might therefore expect corrections of relative order $\sigma/m^2$ 
to be parametrically suppressed by a factor $(\sigma/m^2)^{1/3}$ with respect to corrections 
of relative order $1/(m \langle nl| r |nl\rangle)^2$, if $m \gg \sqrt{\sigma}$. 
Corrections of relative order $\sigma/m^2$ are those associated with the $1/m^2$ potentials $V_{r}^{(2,0)}(r)$ and $V_{r}^{(1,1)}(r)$. 
Corrections of relative order  $1/(m \langle nl| r |nl\rangle)^2$ are those associated with the other $1/m^2$ potentials 
and with the second-order quantum-mechanical corrections induced by the $V^{(1/m)}$ potential.
As we will see, however, for the range of masses considered here, the contributions to the spectrum 
turn out to be numerically comparable for all the $1/m^2$ potentials.
}
\be
E_{nljs} = E^{(0)}_{nl} + \langle nl |V^{(1/m)} |nl\rangle
+ \!\!\sum_{(n',l')\neq (n,l)}\frac{|\langle nl | V^{(1/m)} |n'l'\rangle|^2}{E^{(0)}_{nl}-E^{(0)}_{n'l'}} 
+ \langle nljs| V^{(1/m^2)} |nljs\rangle\,.
\label{ESTlevels}
\ee

\begin{table}[!htbp]
\makebox[6cm]{\phantom b}
\begin{center}
\begin{tabular}{|c||c||c|c|c|c|c||c|}
\hline
Levels & $E^{(0)}$ & $V^{(1/m)}$ & $V^{(1/m)}_{\rm 2nd\, order}$ & $V_{{\bf L}^2}$ & $V_{LS}$  & $V_r$ &  $E$ \\\hline
$1S$ & 1.621 & -0.007 & -0.007 &  0&  0&  -0.021 & 1.586 \\\hline
$1^1P_1$ & 2.331 & 0.080 & -0.005 & -0.027 & 0 &  -0.030 & 2.349 \\\hline
$1^3P_0$ & 2.331 & 0.080 & -0.005 & -0.027 & 0.082  & -0.030 & 2.431 \\\hline
$1^3P_1$ & 2.331 & 0.080 & -0.005 & -0.027 & 0.041  & -0.030 & 2.390 \\\hline
$1^3P_2$ & 2.331 & 0.080 & -0.005 & -0.027 & -0.041 & -0.030 & 2.308 \\\hline
$2S$ & 2.834 & 0.100 & -0.004 & 0 & 0 & -0.037  & 2.893 \\\hline
$1^1D_2$ & 2.946 & 0.134 & -0.004 & -0.062 & 0 & -0.038  & 2.976 \\\hline
$1^3D_1$ & 2.946 & 0.134 & -0.004 & -0.062 & 0.093 & -0.038  & 3.069  \\\hline
$1^3D_2$ & 2.946 & 0.134 & -0.004 & -0.062 & 0.031 & -0.038  & 3.007  \\\hline
$1^3D_3$ & 2.946 & 0.134 & -0.004 & -0.062 & -0.062 & -0.038  & 2.914  \\\hline
$2^1P_1$ & 3.387 & 0.147 & -0.003 & -0.022 &  0  &  -0.044 & 3.465  \\\hline
$2^3P_0$ & 3.387 & 0.147 & -0.003 & -0.022 & 0.066 & -0.044 & 3.531  \\\hline
$2^3P_1$ & 3.387 & 0.147 & -0.003 & -0.022 & 0.033 & -0.044 & 3.498  \\\hline
$2^3P_2$ & 3.387 & 0.147 & -0.003 & -0.022 & -0.033 & -0.044 & 3.432 \\\hline
$3S$ & 3.828 & 0.161 & -0.003 & 0 & 0 & -0.049  & 3.937  \\\hline
$4S$ & 4.706 & 0.203 & -0.002 & 0 & 0 & -0.061  & 4.846  \\\hline
$5S$ & 5.508 & 0.235 & -0.002 & 0 & 0 & -0.071  & 5.670  \\\hline
$6S$ & 6.256 & 0.262 & -0.002 & 0 & 0 & -0.081 & 6.435 \\\hline
\end{tabular}
\end{center}
\caption{Spectrum in the case $m=3\sqrt{\sigma}$. All energies are expressed in units of $\sqrt{\sigma}$.
The column $E^{(0)}$ lists the zeroth-order energy levels, which for $S$ waves are related to the zeros 
of the Airy function~\cite{Lucha:1991vn}.
The column $V^{(1/m)}$ lists the matrix element of $\sigma\ln\left(\sigma r^2\right)/(\pi m)$.
The columns $V^{(1/m)}_{\rm 2nd\, order}$, $V_{{\bf L}^2}$, $V_{LS}$ and $V_r$ list the matrix elements of the 
second-order contribution of the $1/m$ potential and the matrix elements of 
$-\sigma\mathbf{L}^2/(6m^2r)$,  $-\sigma \mathbf{L}\cdot\mathbf{S}/(2m^2r)$ and $-9\,\zeta_3\,\sigma^2r/(2\pi^3m^2)$
respectively. The column $E$ gives the total energy levels according to~\eqref{ESTlevels}.
}
\label{table1}
\end{table}

\begin{table}[htb]
\makebox[6cm]{\phantom b}
\begin{center}
\begin{tabular}{|c||c||c|c|c|c|c||c|}
\hline
Levels & $E^{(0)}$ & $V^{(1/m)}$ & $V^{(1/m)}_{\rm 2nd\, order}$ & $V_{{\bf L}^2}$ & $V_{LS}$  & $V_r$ &  $E$ \\\hline
$1S$ & 1.085 & -0.028 & -0.001 & 0 & 0 & -0.001 & 1.055 \\\hline
$1^1P_1$ & 1.560 & -0.0015 & -0.0007 & -0.004 & 0 &  -0.002 & 1.552  \\\hline
$1^3P_0$ & 1.560 & -0.0015 & -0.0007 & -0.004 & 0.011   & -0.002 &  1.563 \\\hline
$1^3P_1$ & 1.560 & -0.0015 & -0.0007 & -0.004 & 0.006  & -0.002 &  1.558 \\\hline
$1^3P_2$ & 1.560 & -0.0015 & -0.0007 & -0.004 & -0.006 & -0.002 &  1.546 \\\hline
$2S$ & 1.897 & 0.004 & -0.0005 & 0 & 0 & -0.002  & 1.899  \\\hline
$1^1D_2$ & 1.972 & 0.015 & -0.0005 & -0.008 & 0 & -0.002  & 1.977  \\\hline
$1^3D_1$ & 1.972 & 0.015 & -0.0005 & -0.008 & 0.013 & -0.002  & 1.990  \\\hline
$1^3D_2$ & 1.972 & 0.015 & -0.0005 & -0.008 & 0.004 & -0.002  & 1.981 \\\hline
$1^3D_3$ & 1.972 & 0.015 & -0.0005 & -0.008 & -0.008 & -0.002  & 1.969  \\\hline
$2^1P_1$ & 2.267 & 0.019 & -0.0005 & -0.003 & 0 &  -0.003 & 2.280  \\\hline
$2^3P_0$ & 2.267 & 0.019 & -0.0005 & -0.003 & 0.009& -0.003 &2.289  \\\hline
$2^3P_1$ & 2.267 & 0.019 & -0.0005 & -0.003 & 0.004& -0.003 & 2.284 \\\hline
$2^3P_2$ & 2.267 & 0.019 & -0.0005 & -0.003 & -0.004&  -0.003 & 2.276 \\\hline
$3S$ & 2.562 & 0.023 & -0.0004 & 0 & 0 & -0.003  & 2.582  \\\hline
$4S$ & 3.150 & 0.035 & -0.0003 & 0 & 0 & -0.004  & 3.181  \\\hline
$5S$ & 3.687 & 0.045 & -0.0002 & 0 & 0 & -0.004  & 3.728  \\\hline
$6S$ & 4.188 & 0.053 & -0.0002 & 0 & 0 & -0.005 &  4.236\\\hline
\end{tabular}
\end{center}
\caption{Spectrum in the case $m=10\sqrt{\sigma}$, columns are like those in table~\ref{table1}.}
\label{table2}
\end{table}

The results for the spectrum are summarized in the tables~\ref{table1} and~\ref{table2}, which 
refer to the cases $m = 3\sqrt{\sigma}$ and $m=10\sqrt{\sigma}$ respectively.\footnote{
If $\sqrt{\sigma}=457$ MeV~\cite{Koma:2012bc}, then $m = 3\sqrt{\sigma}$ corresponds approximately to the 
charm mass and $m = 10\sqrt{\sigma}$ to the bottom mass.
} 
The tables show all levels up to $n=3$ and all $S$-wave levels up to $n=6$. 
$S$-wave levels are degenerate in spin because the last line of \eqref{model} does not contain a spin-spin interaction.
For some states the $1/m$ potential turns out to give a smaller contribution than the $1/m^2$ potentials. 
It happens when $\sqrt{\sigma} \, \langle nl| r |nl \rangle$ is close to~1, and the 
logarithm in the $1/m$ potential vanishes. This is the case for the $1S$ state 
when $m = 3 \sqrt{\sigma}$: $\sqrt{\sigma} \, \langle 1S| r |1S \rangle \approx 1.08$, 
and for the $1P$ states when $m = 10 \sqrt{\sigma}$: $\sqrt{\sigma} \, \langle 1P| r |1P \rangle \approx 1.04$. 
For the other states and in particular for higher states the contributions of the different 
potentials scale naturally. All $1/m^2$ corrections are of similar size.
This holds also for the newly calculated corrections, which are listed in the column labeled $V_r$, 
showing the relevance of the spin and momentum-independent potentials.

\begin{figure}[!htbp]
\noindent\begin{minipage}[b]{.5\textwidth}
\includegraphics[width=\linewidth]{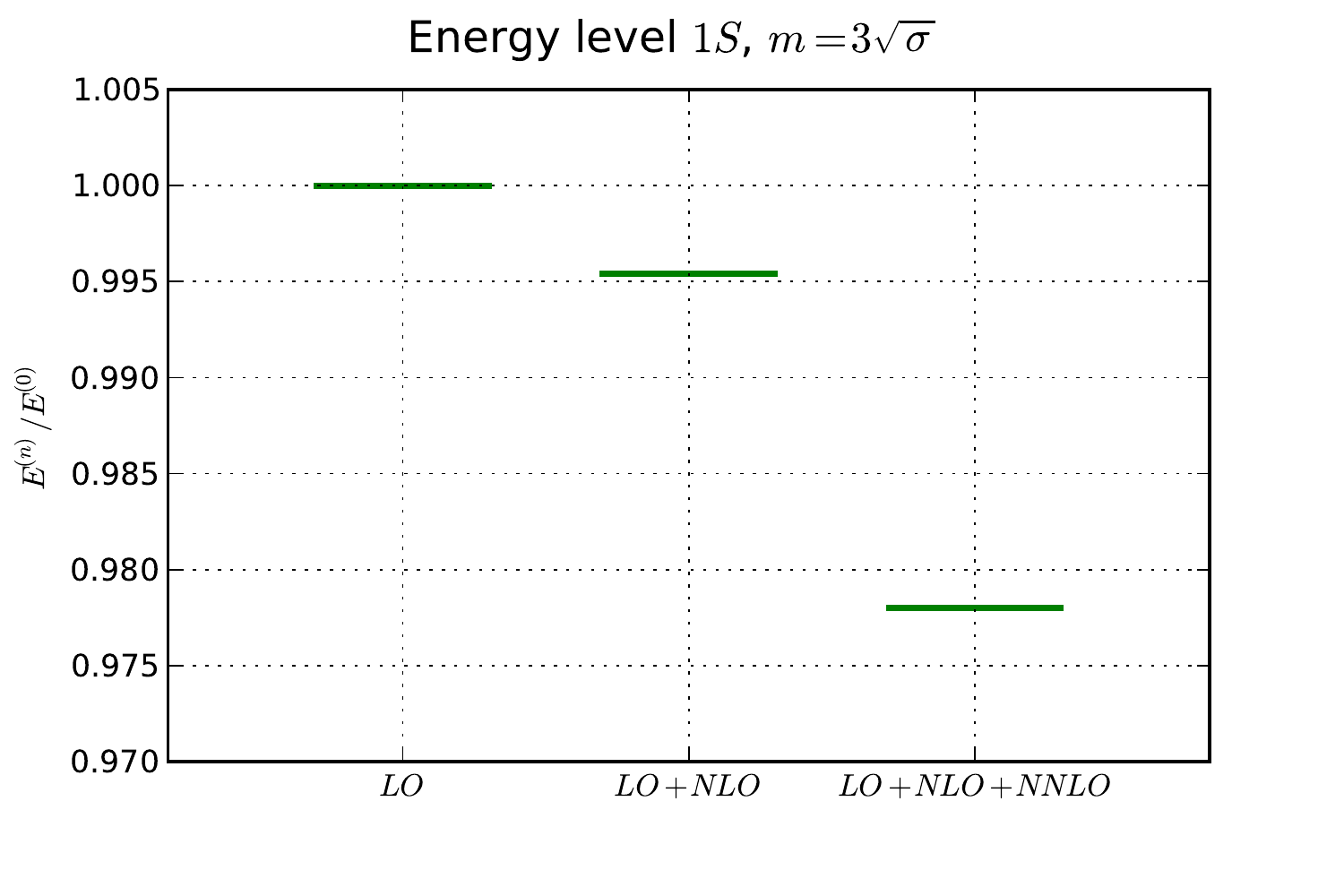}
\end{minipage}%
\hfill
\begin{minipage}[b]{.5\linewidth}
\includegraphics[width=\textwidth]{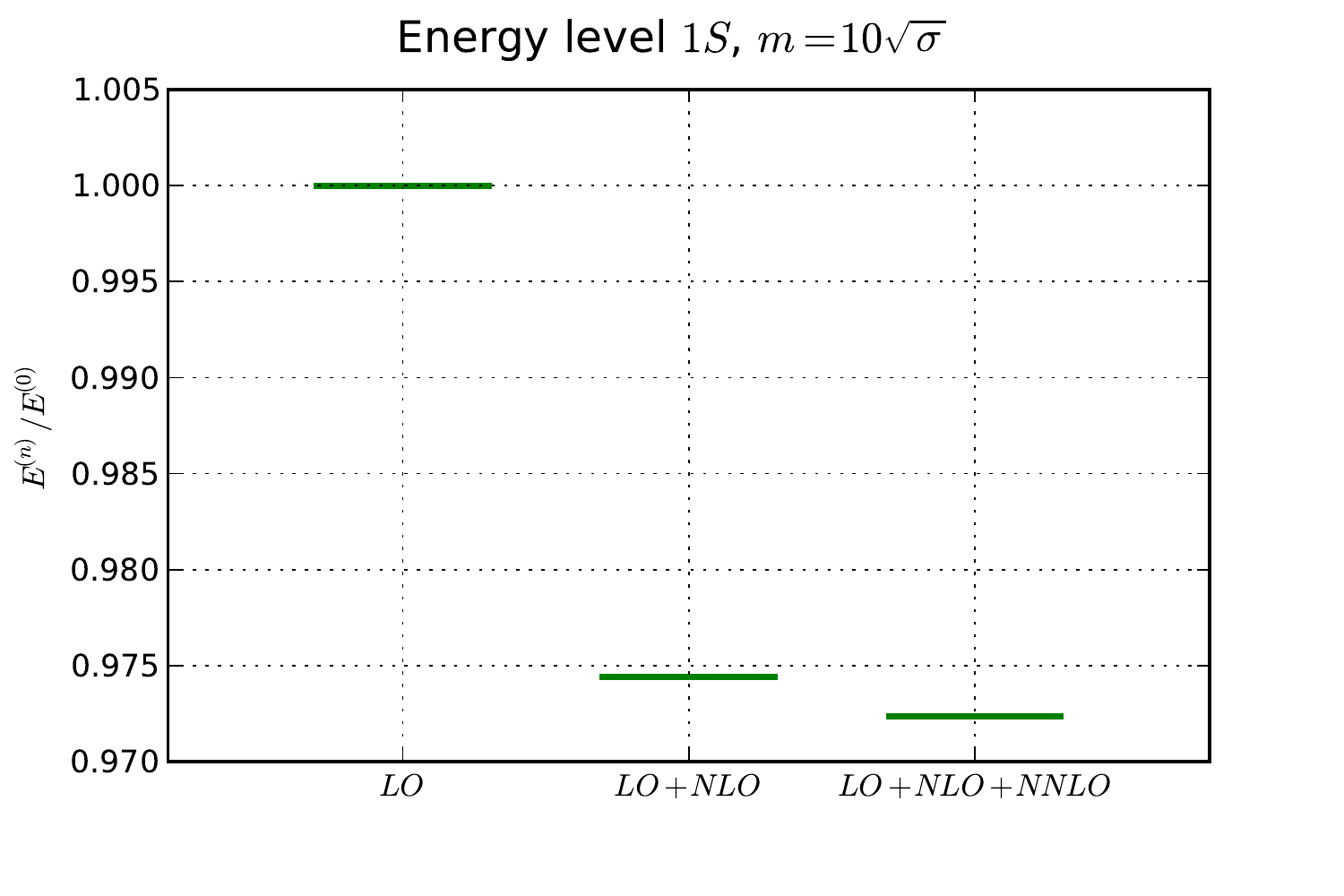}
\par\vspace{0pt}
\end{minipage}
\noindent\begin{minipage}[b]{.5\textwidth}
\includegraphics[width=\linewidth]{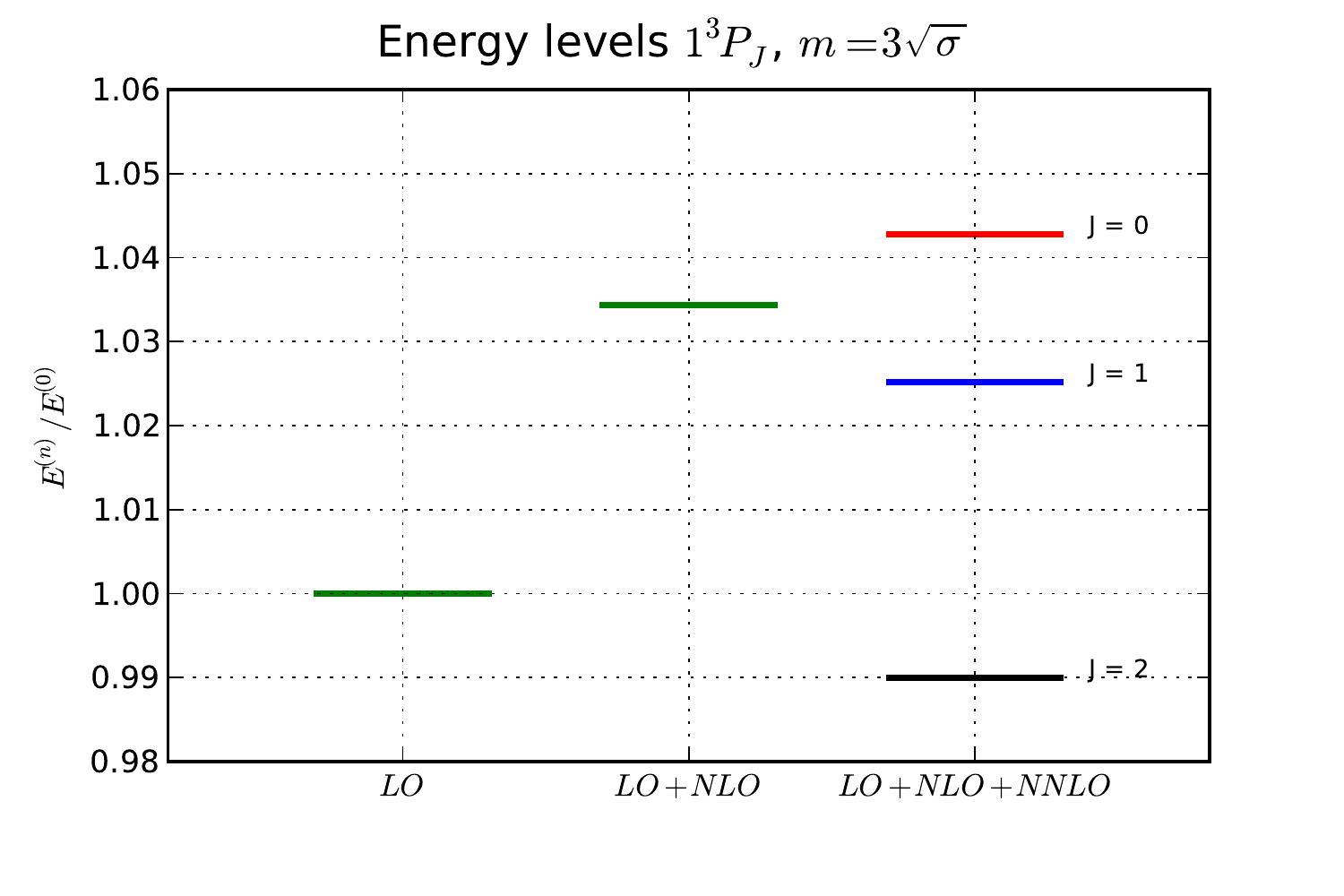}
\end{minipage}%
\hfill
\begin{minipage}[b]{.5\linewidth}
\includegraphics[width=\textwidth]{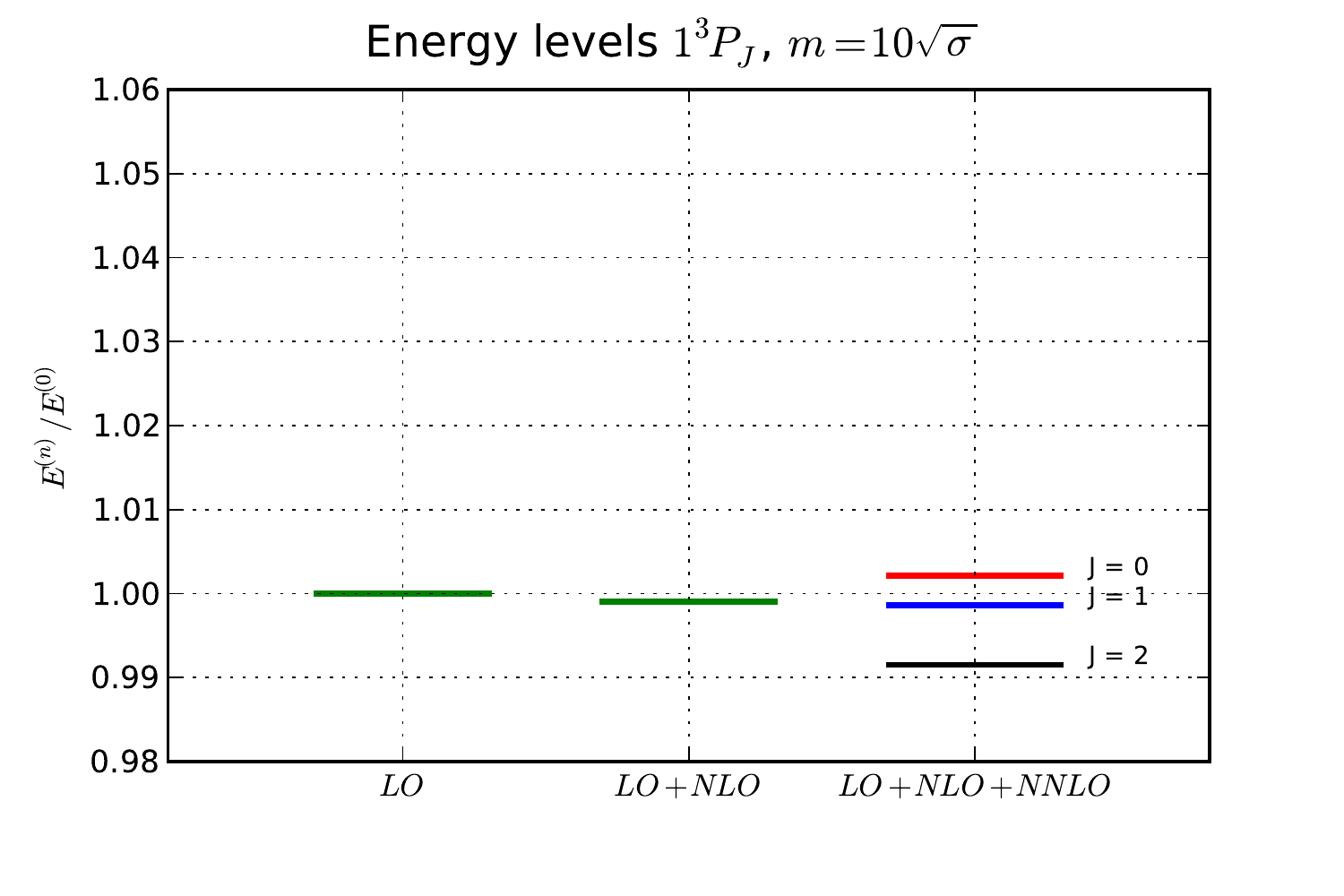}
\end{minipage}
\noindent\begin{minipage}[b]{.5\textwidth}
\includegraphics[width=\linewidth]{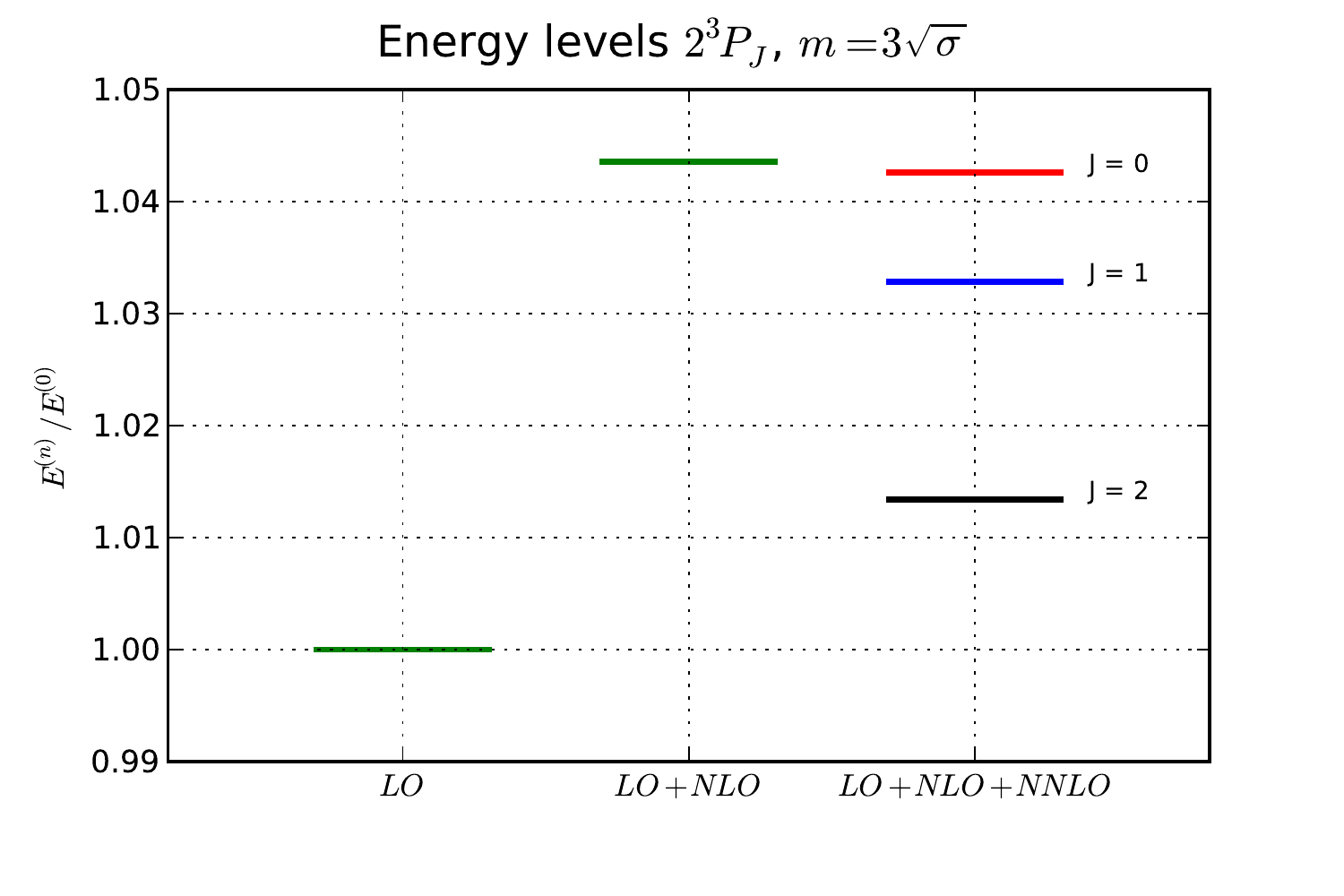}
\end{minipage}%
\hfill
\begin{minipage}[b]{.5\linewidth}
\includegraphics[width=\textwidth]{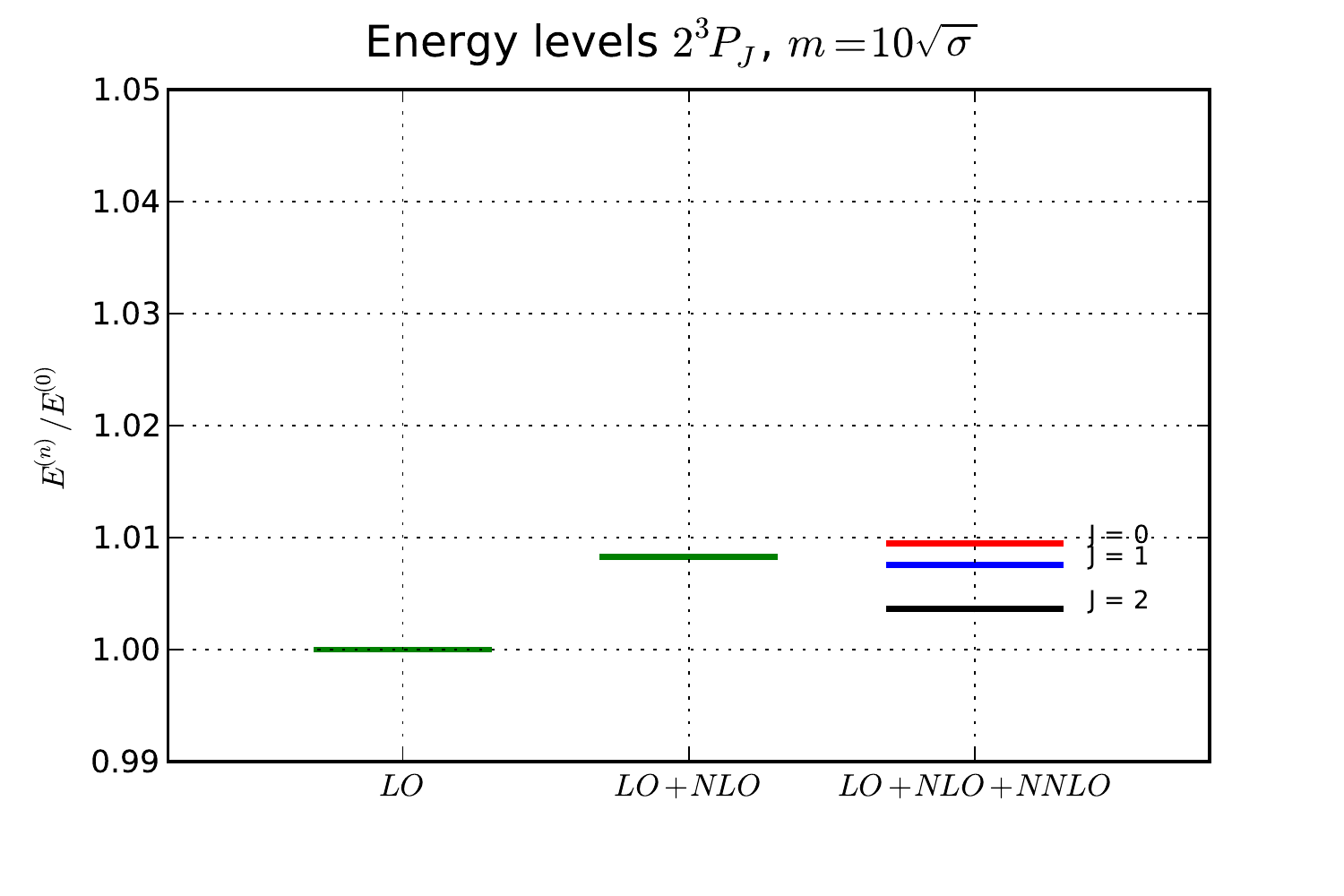}
\end{minipage}
\caption{Energy levels for the states $1S$, $1^3P_J$ and $2^3P_J$ normalized 
with respect to $E^{(0)}_{1S}$,  $E^{(0)}_{1P}$ and  $E^{(0)}_{2P}$ respectively.
The left plots refer to the case $m=3\sqrt{\sigma}$, the right ones to the case $m=10\sqrt{\sigma}$.
The leading order (LO) levels correspond to $E^{(0)}_{nl}$, the next-to-leading-order (NLO) corrections  
to $\langle nl| V^{(1/m)}|nl\rangle$ and the next-to-next-to-leading-order (NNLO) ones 
to the remaining two terms shown in the right-hand side of \eqref{ESTlevels}.}
\label{levels}
\end{figure}

\begin{figure}[!htbp]
\includegraphics[width=0.9\linewidth]{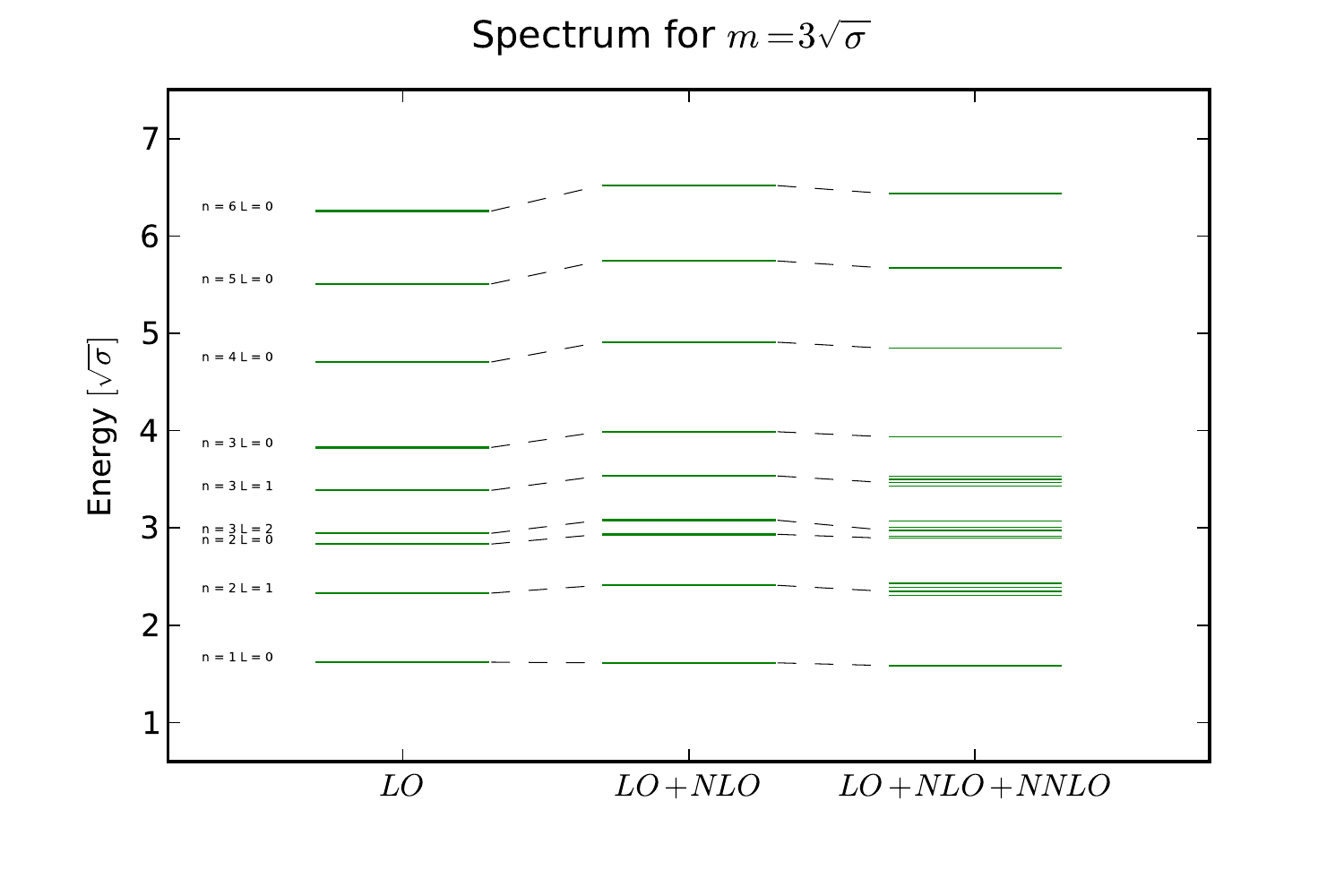}
\caption{Spectrum of all states up to $n=3$ and of all $S$-wave states up to $n=6$ in the case $m=3\sqrt{\sigma}$. 
Energies are expressed in units of $\sqrt{\sigma}$.}
\label{spectrum}
\end{figure}

In figure~\ref{levels} we show graphically the effects of the relativistic corrections to the energy levels  
for the $1S$, $1^3P_J$ and $2^3P_J$ states in the cases $m=3\sqrt{\sigma}$ and $m=10\sqrt{\sigma}$.
In figure~\ref{spectrum} we summarize in one plot the effect of these corrections on the
whole spectrum for the case~$m=3\sqrt{\sigma}$.

\section{Conclusions}
\label{sec_conclusions}
The effective string theory provides an economical way to parameterize the long-range behaviour 
of the heavy quark-antiquark potential in the absence of available lattice data.
Whenever lattice data are available they compare favourably with the EST predictions. 
This is the case for the static potential that has been tested also at the level of  
quantum fluctuations of order $1/r$, the $1/m$ potential, and the $1/m^2$ spin-orbit and 
momentum-dependent potentials. These successful comparisons support the assumption 
of a one-to-one mapping in the long range between Wilson loop expectation values 
and correlators of string coordinates; see~\eqref{mapping}. 

Existing lattice data for the spin-spin potentials are so far consistent 
with zero in the long range~\cite{Koma:2006fw}. It would be interesting 
to produce more accurate data able to detect a long-distance signal, for the EST 
predicts a sharp falloff proportional to $1/r^5$.

In this paper, we have computed in the EST the momentum and spin-independent $1/m^2$ potentials.
They show a linearly rising behaviour with the distance and may be interpreted 
as a sort of relativistic correction to the static potential. 
This is again a sharp prediction of the EST that can be checked against data from lattice, 
once calculations of Wilson loop expectation values with four chromoelectric field insertions are performed.
Under the assumption of the exact mapping~\eqref{mapping} the expressions of the potentials are given in \eqref{v20restlong} and \eqref{v11restlong}.
The net effect of these potentials in the equal mass case is to reduce the string tension by an amount~$9\,\zeta_3\,\sigma^2/(2\pi^3m^2)$.\footnote{
It is interesting to notice that an effective reduction in the string tension due to relativistic effects may be observed 
in some plots of~\cite{Kawanai:2013aca}. We thank Shoichi Sasaki for communications on this point.}

One may argue that the newly computed potentials are of phenomenological relevance in quarkonium physics~\cite{Brambilla:2004wf,Brambilla:2010cs} 
since their contribution to the spectrum, at least when the short-distance part of the potentials is neglected, 
is comparable in size to that of the other $1/m^2$ potentials. 
A realistic description of quarkonium requires, however, the inclusion of the short-distance 
parts of the potentials. These are known from perturbation theory. Spectroscopy studies that
use lattice data to parameterize the long-distance parts of the potentials and perturbation theory for 
the short-distance parts are for instance in~\cite{Bali:1997am,Koma:2012bc,Laschka:2012cf,Pietrulewicz:2013ct}. 
However, such studies are unavoidably incomplete insofar as not all potentials have been computed yet on the lattice.
The core message of this work is that the EST may provide the missing information through 
the long-distance expression of the potential. In the model defined by equation \eqref{model}, 
that expression depends on just two parameters: the heavy-quark mass and the string tension. 
It therefore provides a simple infrared completion of the heavy quark-antiquark potential 
valuable for future quarkonium studies~\cite{Brambilla:2014tr}.

\section*{Acknowledgements}
We thank Guillerm P\'erez-Nadal and Joan Soto for valuable correspondence. 
We are especially grateful to Guillerm P\'erez-Nadal for sharing with us his unpublished notes of Ref.~\cite{PerezNadal:2008vm}.
N.B. thanks Marshall Baker for discussions. 
N.B. and A.V. acknowledge financial support from the DFG cluster of excellence ``Origin and structure of the universe''
(\href{http://www.universe-cluster.de}{www.universe-cluster.de}). H.M. acknowledges financial support from 
the German Academic Exchange Service (DAAD).
This work is supported in part by the Deutsche Forschungsgemeinschaft (DFG) 
and the National Natural Science Foundation (NSFC) through
funds provided to the Sino-German CRC 110 ``Symmetries and
the Emergence of Structure in QCD''.


\begin{thebibliography}{99}
\bibitem{Wilson:1974sk} 
  K.~G.~Wilson,
  Phys.\ Rev.\ D {\bf 10}, 2445 (1974).

\bibitem{Susskind:1976pi} 
  L.~Susskind,
  In *Les Houches 1976, Proceedings, Weak and Electromagnetic Interactions At High Energies*, Amsterdam 1977, 207.

\bibitem{Fischler:1977yf} 
  W.~Fischler,
  Nucl.\ Phys.\ B {\bf 129}, 157 (1977).

\bibitem{Brown:1979ya} 
  L.~S.~Brown and W.~I.~Weisberger,
  Phys.\ Rev.\ D {\bf 20}, 3239 (1979).

\bibitem{Eichten:1980mw} 
  E.~Eichten and F.~Feinberg,
  Phys.\ Rev.\ D {\bf 23}, 2724 (1981).

\bibitem{Peskin:1983up} 
  M.~E.~Peskin,
  In *Proceeding of the 11th SLAC Institute*,  SLAC 1983, 
  Report No. 267, 151. 

\bibitem{Barchielli:1986zs} 
  A.~Barchielli, E.~Montaldi and G.~M.~Prosperi,
  Nucl.\ Phys.\ B {\bf 296}, 625 (1988)
  [Erratum-ibid.\ B {\bf 303}, 752 (1988)].

\bibitem{Brambilla:2000gk} 
  N.~Brambilla, A.~Pineda, J.~Soto and A.~Vairo,
  Phys.\ Rev.\ D {\bf 63}, 014023 (2000)
  [hep-ph/0002250].

\bibitem{Pineda:2000sz} 
  A.~Pineda and A.~Vairo,
  Phys.\ Rev.\ D {\bf 63}, 054007 (2001)
  [Erratum-ibid.\ D {\bf 64}, 039902 (2001)]
  [hep-ph/0009145].

\bibitem{Brambilla:2003mu} 
  N.~Brambilla, A.~Pineda, J.~Soto and A.~Vairo,
  Phys.\ Lett.\ B {\bf 580}, 60 (2004)
  [hep-ph/0307159].

\bibitem{Brambilla:2004jw} 
  N.~Brambilla, A.~Pineda, J.~Soto and A.~Vairo,
  Rev.\ Mod.\ Phys.\  {\bf 77}, 1423 (2005)
  [hep-ph/0410047].

\bibitem{Caswell:1985ui} 
  W.~E.~Caswell and G.~P.~Lepage,
  Phys.\ Lett.\ B {\bf 167}, 437 (1986).

\bibitem{Bodwin:1994jh} 
  G.~T.~Bodwin, E.~Braaten and G.~P.~Lepage,
  Phys.\ Rev.\ D {\bf 51}, 1125 (1995)
  [Erratum-ibid.\ D {\bf 55}, 5853 (1997)]
  [hep-ph/9407339].

\bibitem{Michael:1985wf} 
  C.~Michael and P.~E.~L.~Rakow,
  Nucl.\ Phys.\ B {\bf 256}, 640 (1985).

\bibitem{Michael:1985rh} 
  C.~Michael,
  Phys.\ Rev.\ Lett.\  {\bf 56}, 1219 (1986).

\bibitem{Campostrini:1986ki} 
  M.~Campostrini, K.~Moriarty and C.~Rebbi,
  Phys.\ Rev.\ Lett.\  {\bf 57}, 44 (1986).

\bibitem{Campostrini:1987hu} 
  M.~Campostrini, K.~Moriarty and C.~Rebbi,
  Phys.\ Rev.\ D {\bf 36}, 3450 (1987).

\bibitem{deForcrand:1985zc} 
  P.~de Forcrand and J.~D.~Stack,
  Phys.\ Rev.\ Lett.\  {\bf 55}, 1254 (1985).

\bibitem{Huntley:1986de} 
  A.~Huntley and C.~Michael,
  Nucl.\ Phys.\ B {\bf 286}, 211 (1987).

\bibitem{Koike:1989jf} 
  Y.~Koike,
  Phys.\ Lett.\ B {\bf 216}, 184 (1989).

\bibitem{Born:1993cp} 
  K.~D.~Born, E.~Laermann, T.~F.~Walsh and P.~M.~Zerwas,
  Phys.\ Lett.\ B {\bf 329}, 332 (1994).

\bibitem{Bali:1997am} 
  G.~S.~Bali, K.~Schilling and A.~Wachter,
  Phys.\ Rev.\ D {\bf 56}, 2566 (1997)
  [hep-lat/9703019].

\bibitem{Bali:2000gf} 
  G.~S.~Bali,
  Phys.\ Rept.\  {\bf 343}, 1 (2001)
  [hep-ph/0001312].

\bibitem{Koma:2006si} 
  Y.~Koma, M.~Koma and H.~Wittig,
  Phys.\ Rev.\ Lett.\  {\bf 97}, 122003 (2006)
  [hep-lat/0607009].

\bibitem{Koma:2006fw} 
  Y.~Koma and M.~Koma,
  Nucl.\ Phys.\ B {\bf 769}, 79 (2007)
  [hep-lat/0609078].

\bibitem{Koma:2007jq} 
  Y.~Koma, M.~Koma and H.~Wittig,
  PoS LAT {\bf 2007}, 111 (2007)
  [arXiv:0711.2322 [hep-lat]].

\bibitem{Koma:2009ws} 
  Y.~Koma and M.~Koma,
  PoS LAT {\bf 2009}, 122 (2009)
  [arXiv:0911.3204 [hep-lat]].

\bibitem{Koma:2010zz} 
  Y.~Koma and M.~Koma,
  Prog.\ Theor.\ Phys.\ Suppl.\  {\bf 186}, 205 (2010).

\bibitem{Nambu:1978bd} 
  Y.~Nambu,
  Phys.\ Lett.\ B {\bf 80}, 372 (1979).

\bibitem{Luscher:1980fr} 
  M.~L\"uscher, K.~Symanzik and P.~Weisz,
  Nucl.\ Phys.\ B {\bf 173}, 365 (1980).

\bibitem{Luscher:1980ac} 
  M.~L\"uscher,
  Nucl.\ Phys.\ B {\bf 180}, 317 (1981).

\bibitem{Polchinski:1991ax} 
  J.~Polchinski and A.~Strominger,
  Phys.\ Rev.\ Lett.\  {\bf 67}, 1681 (1991).

\bibitem{Luscher:2002qv} 
  M.~L\"uscher and P.~Weisz,
  JHEP {\bf 0207}, 049 (2002)
  [hep-lat/0207003].

\bibitem{Kogut:1981gm} 
  J.~B.~Kogut and G.~Parisi,
  Phys.\ Rev.\ Lett.\  {\bf 47}, 1089 (1981).

\bibitem{PerezNadal:2008vm} 
  G.~Perez-Nadal and J.~Soto,
  Phys.\ Rev.\ D {\bf 79}, 114002 (2009)
  [arXiv:0811.2762 [hep-ph]].

\bibitem{Barchielli:1988zp} 
  A.~Barchielli, N.~Brambilla and G.~M.~Prosperi,
  Nuovo Cim.\ A {\bf 103}, 59 (1990).

\bibitem{Brambilla:1993zw} 
  N.~Brambilla, P.~Consoli and G.~M.~Prosperi,
  Phys.\ Rev.\ D {\bf 50}, 5878 (1994)
  [hep-th/9401051].

\bibitem{Brambilla:1996aq} 
  N.~Brambilla and A.~Vairo,
  Phys.\ Rev.\ D {\bf 55}, 3974 (1997)
  [hep-ph/9606344].

\bibitem{Brambilla:1999ja} 
  N.~Brambilla and A.~Vairo,
  In *Newport News 1998, Strong interactions at low and intermediate energies* 151-220
  [hep-ph/9904330].

\bibitem{Manohar:1997qy} 
  A.~V.~Manohar,
  Phys.\ Rev.\ D {\bf 56}, 230 (1997)
  [hep-ph/9701294].

\bibitem{Pineda:1998kj} 
  A.~Pineda and J.~Soto,
  Phys.\ Rev.\ D {\bf 58}, 114011 (1998)
  [hep-ph/9802365].

\bibitem{Brambilla:2014jmp} 
  N.~Brambilla, S.~Eidelman, P.~Foka, S.~Gardner, A.~S.~Kronfeld, M.~G.~Alford, R.~Alkofer and M.~Butenschoen {\it et al.},
  arXiv:1404.3723 [hep-ph].

\bibitem{Aharony:2010cx} 
  O.~Aharony and M.~Field,
  JHEP {\bf 1101}, 065 (2011)
  [arXiv:1008.2636 [hep-th]].

\bibitem{Aharony:2013ipa} 
  O.~Aharony and Z.~Komargodski,
  JHEP {\bf 1305}, 118 (2013)
  [arXiv:1302.6257 [hep-th]].

\bibitem{Baker:2000ci} 
  M.~Baker and R.~Steinke,
  Phys.\ Rev.\ D {\bf 63}, 094013 (2001)
  [hep-ph/0006069].

\bibitem{Baker:2002km} 
  M.~Baker and R.~Steinke,
  Phys.\ Rev.\ D {\bf 65}, 094042 (2002)
  [hep-th/0201169].

\bibitem{Luscher:2004ib} 
  M.~L\"uscher and P.~Weisz,
  JHEP {\bf 0407}, 014 (2004)
  [hep-th/0406205].

\bibitem{Martinez:2012rra} 
  H.~E.~Martinez,
  PoS ConfinementX, 161 (2012).

\bibitem{Brambilla:2001xk} 
  N.~Brambilla, D.~Gromes and A.~Vairo,
  Phys.\ Rev.\ D {\bf 64}, 076010 (2001)
  [hep-ph/0104068].

\bibitem{Brambilla:2003nt} 
  N.~Brambilla, D.~Gromes and A.~Vairo,
  Phys.\ Lett.\ B {\bf 576}, 314 (2003)
  [hep-ph/0306107].

\bibitem{Gromes:1984ma} 
  D.~Gromes,
  Z.\ Phys.\ C {\bf 26}, 401 (1984).

\bibitem{Lucha:1991vn} 
  W.~Lucha, F.~F.~Sch\"oberl and D.~Gromes,
  Phys.\ Rept.\  {\bf 200}, 127 (1991).

\bibitem{Koma:2012bc} 
  Y.~Koma and M.~Koma,
  PoS LATTICE {\bf 2012}, 140 (2012)
  [arXiv:1211.6795 [hep-lat]].

\bibitem{Kawanai:2013aca} 
  T.~Kawanai and S.~Sasaki,
  Phys.\ Rev.\ D {\bf 89}, 054507 (2014)
  [arXiv:1311.1253 [hep-lat]].

\bibitem{Brambilla:2004wf}
  N.~Brambilla {\it et al.},
  CERN-2005-005, (CERN, Geneva, 2005)
  [arXiv:hep-ph/0412158].

\bibitem{Brambilla:2010cs} 
  N.~Brambilla, S.~Eidelman, B.~K.~Heltsley, R.~Vogt, G.~T.~Bodwin, E.~Eichten, A.~D.~Frawley and A.~B.~Meyer {\it et al.},
  Eur.\ Phys.\ J.\ C {\bf 71}, 1534 (2011)
  [arXiv:1010.5827 [hep-ph]].

\bibitem{Laschka:2012cf} 
  A.~Laschka, N.~Kaiser and W.~Weise,
  Phys.\ Lett.\ B {\bf 715}, 190 (2012)
  [arXiv:1205.3390 [hep-ph]].

\bibitem{Pietrulewicz:2013ct} 
  P.~Pietrulewicz,
  PoS ConfinementX, 135 (2012)
  [arXiv:1301.1308 [hep-ph]].

\bibitem{Brambilla:2014tr}
  N.~Brambilla, H.~E.~Martinez and A.~Vairo, 
  TUM-EFT 40/13, in preparation.  

\end{thebibliography}
\end{document}